\documentclass[iop, twocolumn]{aastex63}
\hypersetup{citecolor=blue,urlcolor=red}
\shorttitle{PKS 1510$-$089}
\shortauthors{Yuan et al.}
\usepackage{multirow}
\usepackage{soul}
\usepackage{xspace}
\usepackage{subfiles}
\usepackage{ulem}
\usepackage{amssymb}
\usepackage{amsmath}
\usepackage{multirow}
\usepackage{bm}


\begin{document}

\title{Multi-wavelength temporal variability of the blazar PKS 1510$-$089}

\correspondingauthor{Q. Yuan}
\email{yuanqi@xao.ac.cn}

\author[0000-0003-4671-1740]{Q. Yuan}
\affiliation{Xinjiang Astronomical Observatory, Chinese Academy of Sciences, 150 Science 1-Street, Urumqi 830011, China}
\affiliation{University of Chinese Academy of Sciences, 19A Yuquan Road, Beijing 100049, China}

\author[0000-0001-6890-2236]{Pankaj Kushwaha}
\altaffiliation{DST-INSPIRE Faculty}
\affiliation{Department of Physical Sciences, Indian Institute of Science Education and Research Mohali, Knowledge City, Sector 81, SAS Nagar, Punjab 140306, India}
\affiliation{Aryabhatta Research Institute of Observational Sciences (ARIES), Manora Peak, Nainital 263 001, India}

\author[0000-0002-9331-4388]{Alok C. Gupta}
\affiliation{Key Laboratory for Research in Galaxies and Cosmology, Shanghai Astronomical Observatory, Chinese Academy of Sciences, 80 Nandan Road, Shanghai 200030, China}
\affiliation{Aryabhatta Research Institute of Observational Sciences (ARIES), Manora Peak, Nainital 263 001, India}

\author[0000-0002-3960-5870]{Ashutosh Tripathi}
\affiliation{Department of Physics, Southern Methodist University, 3215 Daniel Ave., Dallas, TX 75205, USA}
\affiliation{Center for Field Theory and Particle Physics and Department of Physics, Fudan University, 2005 Songhu Road, Shanghai 200438, China}

\author[0000-0002-1029-3746]{Paul J. Wiita}
\affiliation{Department of Physics, The College of New Jersey, 2000 Pennington Rd., Ewing, NJ 08628-0718, USA}

\author[0000-0002-8315-2848]{M. Zhang} 
\affiliation{Xinjiang Astronomical Observatory, Chinese Academy of Sciences, 150 Science 1-Street, Urumqi 830011, China}
\affiliation{Key Laboratory for Radio Astronomy, Chinese Academy of Sciences, 2 West Beijing Road, Nanjing 210008, China}
\affiliation{Xinjiang Key Laboratory of Radio Astrophysics, 150 Science 1-Street, Urumqi 830011, China}

\author{X. Liu} 
\affiliation{Xinjiang Astronomical Observatory, Chinese Academy of Sciences, 150 Science 1-Street, Urumqi 830011, China}
\affiliation{Key Laboratory for Radio Astronomy, Chinese Academy of Sciences, 2 West Beijing Road, Nanjing 210008, China}
\affiliation{Xinjiang Key Laboratory of Radio Astrophysics, 150 Science 1-Street, Urumqi 830011, China}

\author[0000-0002-0393-0647]{Anne L\"ahteenm\"aki} 
\affiliation{Aalto University Mets\"ahovi Radio Observatory, Mets\"ahovintie 114, 02540 Kylm\"al\"a, Finland}
\affiliation{Aalto University Department of Electronics and Nanoengineering, P.O. BOX 15500, FI-00076 Aalto, Finland}

\author[0000-0003-1249-6026]{Merja Tornikoski} 
\affiliation{Aalto University Mets\"ahovi Radio Observatory, Mets\"ahovintie 114, 02540 Kylm\"al\"a, Finland}

\author[0000-0002-9164-2695]{Joni Tammi} 
\affiliation{Aalto University Mets\"ahovi Radio Observatory, Mets\"ahovintie 114, 02540 Kylm\"al\"a, Finland}

\author[0000-0002-9248-086X]{Venkatessh Ramakrishnan}
\affiliation{Aalto University Mets\"ahovi Radio Observatory, Mets\"ahovintie 114, 02540 Kylm\"al\"a, Finland}
\affiliation{Finnish Centre for Astronomy with ESO (FINCA), University of Turku, Vesilinnantie 5, 20014 University of Turku, Finland}

\author{L. Cui}
\affiliation{Xinjiang Astronomical Observatory, Chinese Academy of Sciences, 150 Science 1-Street, Urumqi 830011, China}
\affiliation{Key Laboratory for Radio Astronomy, Chinese Academy of Sciences, 2 West Beijing Road, Nanjing 210008, China}
\affiliation{Xinjiang Key Laboratory of Radio Astrophysics, 150 Science 1-Street, Urumqi 830011, China}

\author{X. Wang}
\affiliation{Xinjiang Astronomical Observatory, Chinese Academy of Sciences, 150 Science 1-Street, Urumqi 830011, China}
\affiliation{University of Chinese Academy of Sciences, 19A Yuquan Road, Beijing 100049, China}

\author[0000-0002-4455-6946]{M.F. Gu}
\affiliation{Key Laboratory for Research in Galaxies and Cosmology, Shanghai Astronomical Observatory, Chinese Academy of Sciences, 80 Nandan Road, Shanghai 200030, China}

\author[0000-0002-3180-9502]{Cosimo Bambi}
\affiliation{Center for Field Theory and Particle Physics and Department of Physics, Fudan University, 2005 Songhu Road, Shanghai 200438, China}

\author[0000-0002-3839-3466]{A. E. Volvach}
\affiliation{Radio Astronomy and Geodynamics Department CrAO, Taras Shevchenko National University of Kyiv, Ukraine}

\begin{abstract}
\noindent
We perform correlation and periodicity search analyses on long-term multi-band light curves of the FSRQ 1510$-$089 observed by the space-based Fermi--Large Area Telescope in $\gamma$-rays, the SMARTS and Steward Observatory telescopes in optical and near-infrared (NIR) and the 13.7 m radio telescope in Mets\"ahovi Radio Observatory between 2008 and 2018.
The z-transform discrete correlation function method is applied to study the correlation and possible time lags among these multi-band light curves. 
Among all pairs of wavelengths, the $\gamma$-ray vs.\ optical/NIR and optical vs.\ NIR correlations show zero time lags;
however, both the $\gamma$-ray and optical/NIR emissions precede the radio radiation.
The Generalized Lomb-Scargle periodogram, Weighted Wavelet Z-transform, and REDFIT techniques are employed to investigate the unresolved-core-emission dominated 37 GHz light curve and yield evidence for a quasi-period around 1540 days, although given the length of the whole data set it cannot be claimed to be significant.
We also investigate the optical/NIR color variability and find that this source shows a simple redder-when-brighter behavior over time, even in the low flux state.
\end{abstract}


\section{Introduction}
\noindent
Blazars are radio-loud (RL) active galactic nuclei (AGNs) whose relativistic jets are seen at a small angle to the line of sight~\citep{urry.95.pasp}. 
Traditionally, blazars are subclassified as flat spectrum radio quasars (FSRQs) and BL Lac objects (BL Lacs). 
There are strong emission lines in the spectrum of the former while these lines are weak or absent in the latter.
The fluxes from blazars are highly variable in the entire accessible electromagnetic spectrum, with time scales ranging from minutes to years~\citep[e.g.,][and references therein]{1993ApJ...411..614U,2000ApJ...536..742P,2001A&A...367..809K,2001ApJ...559..187K,2002A&A...393...89A,2005A&A...430..865A,2005ApJ...630..130B,2005A&A...442..895A,2006ApJ...641..740R,2007ApJ...664L..71A,2008ApJ...677..906F,2009ApJ...695..596H,2009A&A...502..749A,2017MNRAS.472..788G,2019AJ....157...95G,2022ApJS..260...39G}. \\  
\\
The Spectral Energy Distributions (SEDs) of blazars are characterized by a broad double-peaked structure.
The low-energy hump spans radio to ultraviolet (UV) or X-ray bands with a peak between near-infrared (NIR) and UV/X-ray.
This low-energy hump is synchrotron radiation from the relativistic non-thermal electrons in the jet.
The high-energy hump, on the other hand, extends from X-rays to GeV/TeV $\gamma$-rays, peaking between hard X-rays and $\gamma$-rays.
The origin of the second hump is still unclear; both lepton and hadron based emission scenarios have been proposed~\citep[e.g.,][]{bottcher.13.apj}. 
In leptonic scenarios, the mechanism of the high energy emission is the inverse-Compton scattering of low-energy seed photons by the same relativistic electrons 
that produces the synchrotron radiation.
Those low-energy seed photons may be the synchrotron photons produced in the jet (synchrotron self-Compton, SSC) or be external photons originating in the local environment (external Compton, EC) which includes the accretion disk, the broad line region (BLR) and the dusty torus (DT) or conceivably even the cosmic microwave background \citep[e.g.,][and references therein]{2000ApJ...544L..23T,2013MNRAS.433.2380K}. 
In the hadronic scenario, the second hump is due to the proton synchrotron or the proton-induced particle cascades~\citep[e.g.,][]{mannheim.92.aa,mucke.01.mn}. \\
\\
Though broadly stochastic in nature, flux variability in the various wavebands differs from source to source.
Even for the same source, there can be significant differences in the correlation of light curves between multiple bands during different observation periods, 
so studying multi-band spectral and temporal variations offer a non-direct way to peek into unresolved processes and spatial scales hidden in compact systems. 
In general, the emission variability of blazars may caused by a combination of factors internal to the jet, such as shocks, turbulence, or magnetic reconnection, and external ones, such as changes in the jet direction with respect to the observer~\citep[e.g.][]{bottcher.19.galaxies}. Blazar variability patterns pose challenges to some existing blazar radiation models and can help discriminate between them.\\
\\
The location of the $\gamma$-ray emission site(s) in blazars has been the subject of substantial debate \citep[e.g.][and references therein]{kramarenko.22.mn,2021MNRAS.500.5297A,2018A&A...616A..63A,2018MNRAS.477.4749C,2017A&A...597A..80H,2013MNRAS.433.2380K}.
It is well known that the optical emission of blazars is usually dominated by relativistic jets, but there are exceptions.
For example,~\citet{fernandes.20.mn} found that the optical emission in the FRSQ 3C~273 was dominated by the accretion disk rather than the jet over the entire time span of their study.
A detailed multi-wavelength cross-correlation study can analyze the radiation mechanisms in different bands and provide constraints on the location of the $\gamma$-ray emission region(s). \\
\\
PKS 1510$-$089 is a bright FSRQ located at redshift, $z = 0.36$~\citep{thompson.90.pasp, tanner.96.aj}. 
It is one of the nine FSRQs detected in the very-high-energy (VHE $\ge$ 100 Gev) range by the Imaging Atmospheric Cherenkov Telescopes (IACTs)\footnote{\url{http://tevcat.uchicago.edu}} and also one of the persistent MeV-GeV emitters \citep[e.g.][]{2017ApJ...849..138K}. 
The High Energy Spectroscopic System (H.E.S.S.) collaboration discovered VHE $\gamma-$ray emission $>$ 0.1 TeV from PKS 1510$-$089 in its 15.8 hours of observation (at 9.2$\sigma$)\citep{2013A&A...554A.107H}. 
The source emitted persistent VHE $\gamma-$ray emission in MAGIC observations during its low flux state for an extended period of time during 2012 -- 2017 \citep{2018A&A...619A.159M}. 
PKS 1510$-$089 has witnessed several simultaneous multi-wavelength observing campaigns that obtained flux and spectral variabilities, performed SED modeling, and discussed the diverse emission mechanisms at different epochs of observations \citep[e.g.,][and references therein]{2017A&A...603A..29A,2014A&A...569A..46A,2013MNRAS.428.2418O,2012MNRAS.424..789C,abdo.10.apj,2010ApJ...710L.126M,2008ApJ...672..787K}. 
Recently, transient $\gamma-$ray quasi-periodic oscillations (QPOs) with periods of 3.6 days and 92 days were reported in its Fermi-LAT observations covering $\sim$ 12 years~\citep{2022MNRAS.510.3641R}.\\
\\
PKS 1510$-$089 has had a substantial amount of well-synchronized sampled data taken in optical and NIR bands.
It is also a target source for numerous blazar radio monitoring programs, such as the Fermi-GST AGN Multi-frequency Monitoring Alliance (F-GAMMA) program~\citep{fuhrmann.16.aa,angelakis.19.aa}, the ongoing Owens Valley Radio Observatory (OVRO) 40m monitoring program~\citep{richards.11.apjs}, and the Mets\"ahovi Radio Observatory blazar monitoring program.
In addition to these monitoring campaigns, there are quite a few higher-resolution very-long-baseline interferometry (VLBI) observations: 
the VLBA (Very Long Baseline Array) 2 cm Survey~\citep{kellermann.98.aj} and its successor -- MOJAVE (Monitoring Of Jets in Active galactic nuclei with VLBA Experiments) program~\citep{lister.18.apjs}; the VLBA-BU-BLAZAR program and its successor -- BEAM-ME (Blazars Entering the Astrophysical Multi-Messenger Era) program~\citep{jorstad.16.galaxiesJ,jorstad.17.apj,weaver.22.apjs}. The correlations between different bands have already been studied for PKS 1510$-$089.
\citet{abdo.10.apj} reported that the $\gamma$-ray light curve had a complicated correlation with other wavelengths during the period between September 2008 and June 2009. 
They saw no correlation with the X-ray band, a weak correlation with the UV band, and a significant correlation with the optical band, with the $\gamma$-ray light curve preceding the optical one by about 13 days.
In observations taken during 2009--2013,\citet{beaklini.17.aa} detected a correlation between the radio and $\gamma$-ray flares where the radio flares are delayed by approximately 54 days relative to the $\gamma$-ray flares.
In another study based on observations taken during July 2012 to October 2014, the optical and $\gamma$-ray bands showed correlations with zero time delay ~\citep{ramakrishnan.16.mn}. \\
\\
Broadband SED studies using multi-wavelength data at different source activity states
e.g. August 2006 \citep{kataoka.08.apj}, August 2008 to May 2012 \citep{brown.13.mn}, and very high energy $\gamma-$ray flares in March 2009 \citep{barnacka.14.aap} indicate that the radio and optical/NIR emission is synchrotron radiation from the jet, while the high energy emission (X-ray and $\gamma$-ray) arises from the EC scattering of seed photos from the BLR and DT.\\ 
\\
This paper focuses on multi-band emission connections over lengthy timescales, as we try to better understand blazar radiation mechanisms and jet kinematic behaviors.
Details of the multi-band data we include and their reduction procedures for a ten-year-long observation period are in Section 2, with a study of spectral variations given in Section 2.5.
In section 3, we present the correlation analysis method and report results among different bands.
The subsequent section deals with the methods used to search for QPOs and the results.
We discuss these results in Section 5 and summarize our findings in Section 6.

\section{MULTI-WAVELENGTH ARCHIVAL DATA AND REDUCTION}

\subsection{Fermi $\gamma$-ray data}
\begin{figure*}
    \centering
    \includegraphics[width=18cm]{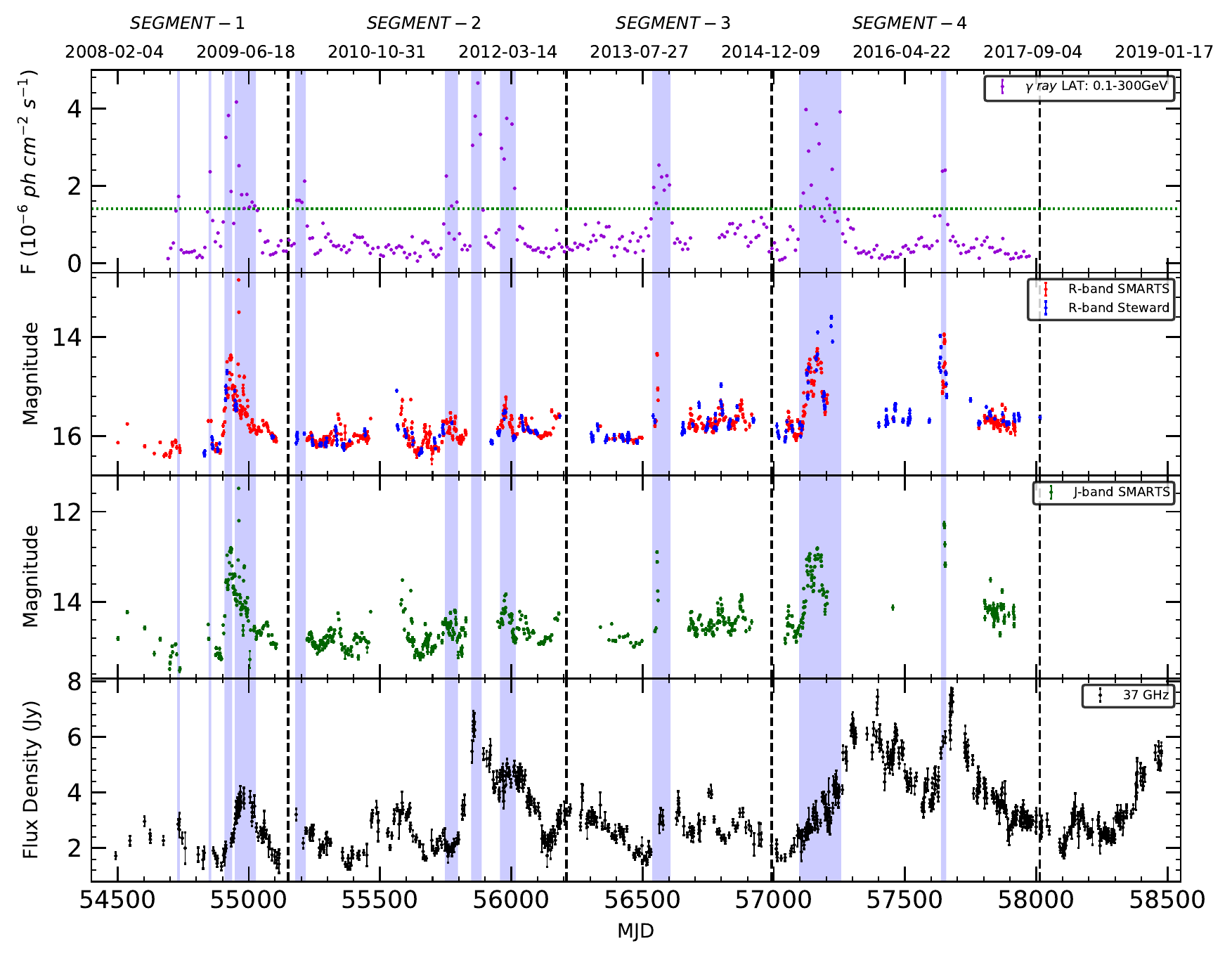}
    \caption{Multiband light curves of PKS 1510$-$089 from $\gamma$-ray to radio.
    Vertical lines divide the segments considered for temporal analysis in our work, the purple-shaded regions represent the $\gamma$-ray flaring states.
    The green horizontal dotted line indicates the constant flux value that divides the high and low $\gamma$-ray states.}
    \label{fig:lc}
\end{figure*}

\noindent
The {\it Fermi} mission is a space-based $\gamma$-ray observatory launched in 2008 \citep{abdo.10.apj}. 
It is sensitive to gamma-ray photons with energies from $\gtrsim 20$ MeV to $\sim$ TeV. 
It continuously scans the sky for $\gamma$-rays and covers the entire sky region within 90 minutes. \\
\\
We analyzed the $\gamma$-ray data from the Large Area Telescope (LAT) following the standard analysis procedures employing a ``binned likelihood" analysis with the {\it Fermipy} (v1.0.1) software.
We downloaded the PASS8 (P8R3) instrument response function to process data from Fermi-LAT. 
First, we applied the intended cuts as recommended by selecting
only the events marked as ``SOURCE" class (evclass=128 and evtype=3) with energies above 100 MeV from a region of interest of $15^\circ$ centered on the source location ($\alpha_{2000.0} =$ 15h 12m 52.2s, $\delta_{2000.0} = -09^\circ \ 06^\prime \ 21.6^{\prime \prime}$). A zenith angle cut of $90^\circ$ was also applied as recommended by the LAT team. 
The good time intervals (GTIs) were generated using the standard criteria "$\rm (DATA\_QUAL \> 0)\&\&(LAT\_CONFIG==1)$". For spectral modeling, an XML file with sources from the fourth LAT point source catalog \citep[4FGL;][]{abdollahi.20.apjs} was used. 
Apart from the $\gamma$-ray emitting point sources, the XML model file also included the galactic and extragalactic isotropic contributions through the template file ``1gll\_psc\_v21.fits" and ``iso\_P8R3\_SOURCE\_V2\_v1.txt", provided by the LAT team.
The source was modeled with a log-parabola spectrum, as in the catalog. 
A test-statistic (TS) value of $\geq 9$ was used to select the data for our analysis. 
The light curve (photon flux vs time) extracted following this procedure is shown in the top panel of Fig.~\ref{fig:lc}.

\subsection{Optical and NIR data}
\noindent
The optical R-band data and NIR J-band data for PKS 1510$-$089 are taken from the public archives of the Small and Moderate Aperture Research Telescope System (SMARTS)\footnote{\url{http://www.astro.yale.edu/smarts/glast/home.php}} and the Steward Observatory telescopes.
SMARTS consists of four meter-class telescopes at the Cerro Tololo Inter-American Observatory (CTIO) in Chile: the 0.9~m, 1.0~m, 1.3~m, and 1.5~m, which observed the Fermi-LAT monitored blazars at both optical and NIR wavelengths.
Details about the SMARTS telescopes, detectors, observations, and data analysis are given in~\citet{bonning.12.apj} and \citet{buxton.12.aj}.\\
\\
The Steward Observatory of the University of Arizona uses the 2.3 m Bok and 1.54 m Kuiper telescopes to carry out optical photometric and polarimetric observations of a large number of blazars using the spectropolarimeter (SPOL).
Details about these telescopes, instruments, observations, and data analysis methods are given in~\citet{smith.09.arxiv}. 
These R-band observations are combined with those of SMARTS in the second panel of Fig.\ \ref{fig:lc} and the SMARTS J-band data are in the third panel. 
The Galactic dust reddening and extinction for a line of sight must be accounted for.
The total Galactic visual extinction is estimated and corrected using the extinction map given by \citet{1998ApJ...500..525S}. 
Newer estimates of Galactic dust extinction from \citet{2011ApJ...737..103S} are now provided alongside those of \citet{1998ApJ...500..525S}. 
The magnitudes are corrected for galactic extinction using an online tool\footnote{\url{https://ned.ipac.caltech.edu/extinction_calculator}} based on~\citet{2011ApJ...737..103S}.

\subsection{Radio Data}
\noindent
The 37 GHz observations of the blazar PKS 1510$-$089 were made with the 13.7 m diameter Aalto University Mets\"ahovi radio telescope, 
which is a radome enclosed Cassegrain type antenna situated in Finland (60$^{\circ}$ 13$^{'}$ 04$^{''}$ N, \ 24$^{\circ}$ 23$^{'}$ 35$^{''}$ E). 
The receivers have HEMPT (high electron mobility pseudomorphic transistor) front ends operating at room temperature. 
The observations are Dicke switched ON--ON observations, alternating the source and the sky in each feed horn. 
A typical integration time to obtain one flux density data point is between 1200 and 1800 s. 
The detection limit of the telescope at 37 GHz is on the order of 0.2 Jy under optimal conditions, but is heavily weather dependent. 
Data points with a signal-to-noise ratio $<$ 4 are handled as non-detections. 
The flux density scale is set by observations of the H II region DR 21. Sources NGC 7027, 3C 274 and 3C 84 are used as secondary calibrators. 
A detailed description of the data reduction and analysis is given in \citet{1998A&AS..132..305T}. 
The error estimate in the flux density includes the contribution from the measurement $RMS$ and the uncertainty of the absolute calibration. 
Radio data at 37~GHz are plotted in the bottom panel of Fig.\ \ref{fig:lc}.

\subsection{Multi-wavelength light curves}
\noindent
Fig.\ref{fig:lc} shows the multi-wavelength light curves of PKS 1510$-$089 between 2008 and 2018.
The aggregated data show frequent changes in the fluxes in different electromagnetic (EM) bands, with the  high (active or flaring) states of the $\gamma$-ray fluxes shaded in purple (Fig.~\ref{fig:lc}); that shading is extended to the other bands for convenient visualization.   To examine the apparent correlations between variations at different wavelengths in detail, we divide the complete data into four individual segments, denoted by vertical dashed lines, each of which has at least one high $\gamma$-ray state 
and concludes in a gap in the optical/NIR corresponding to the end of an observing season. These segments are chosen to be long enough to allow for detection of possible changes in correlations between different bands, as seen in our earlier work \citep[e.g.,][]{2014ApJ...781L...4G,2017MNRAS.472..788G,2017MNRAS.464.2046K,2018MNRAS.473.1145K,2018MNRAS.479.1672K}. \\

\begin{table*}
   \centering
   \caption{Start and Stop dates of the four individual segments}
   \label{tab:seg}
   \begin{tabular}{rrrrr}
      \hline
        SEGMENT & Start Date & Stop Date  & Start MJD & Stop MJD  \\
      \hline
        
        SEGMENT 1  &  2008 Jan 28 &  2009 Oct 11  &  54493 & 55115    \\
        SEGMENT 2  &  2009 Oct 18 &  2012 Oct 02  &  55122 & 56202    \\
        SEGMENT 3  &  2012 Oct 02 &  2014 Dec 01  &  56202 & 56992    \\
        SEGMENT 4  &  2014 Dec 01 &  2017 Sep 18  &  56992 & 58014    \\
     \hline
   \end{tabular}
\end{table*}

\begin{table*}
\centering
\caption{Each sub-segment is defined as the start and the end times of the high state on $\gamma$-ray emission}
\label{tab:flare}
\begin{tabular}{llllll}
\hline
                              &            & Start Date   & Stop Date     & Start MJD   & Stop MJD           \\ \hline
                              & SEGMENT 1-1 & 2008 Sep 18     &  2008 Sep 28    &  54727  &  54737       \\
SEGMENT 1                     & SEGMENT 1-2 & 2009 Jan 16    & 2009 Feb 16     &  54847  &  54857       \\
                              & SEGMENT 1-3 & 2009 Mar 17    & 2009 Apr 16     &  54907  &  54937       \\
                              & SEGMENT 1-4 & 2009 Apr 26    & 2009 Jul 15     &  54947  &  55027       \\ \hline
                              & SEGMENT 2-1 &  2009 Dec 12    &  2010 Jan 21     &  55177  &  55217       \\ 
SEGMENT 2                     & SEGMENT 2-2 &  2011 Jul 5     &  2011 Aug 24     &  55747  &  55797       \\
                              & SEGMENT 2-3 &  2011 Oct 13    &  2011 Nov 22     &  55847  &  55887       \\
                              & SEGMENT 2-4 &  2012 Jan 31    &  2012 Mar 31     &  55957  &  56017       \\ \hline
SEGMENT 3                     & SEGMENT 3-1 &  2013 Sep 2    &  2013 Nov 11      &  56537  & 56607       \\ \hline
SEGMENT 4                     & SEGMENT 4-1 &  2015 Mar 16   &  2015 Aug 23      & 57097  &  57257       \\
                              & SEGMENT 4-2 &  2016 Sep 6    &  2016 Sep 26      & 57637  &  57657       \\ \hline
\end{tabular}
\end{table*}

\noindent
The start and stop dates of the individual segments are given in Table ~\ref{tab:seg}.
The flaring states' start and end times within each segment define sub-segments that are listed in Table ~\ref{tab:flare}.
There are a variety of ways to decide which data points belong to the high and low (quiescent) states~\citep{prince.19.apj, meyer.19.apj} (e.g., fractional rms variability, Bayesian blocks, etc).
Here we follow the work of \citet{kushwaha.16.apjl} who showed that the $\gamma$-ray flux distribution of PKS 1510$-$089 showed two distinct log-normal profiles, one for the high flux level and one for the low one.  The intersection of these two log-normal fits, at $\rm 10^{-5.853} = 1.401\times10^{-6} \ ph\ cm^{-2}\ s^{-1}$, is taken as the separation between flux levels and is shown as 
a green horizontal dotted line running across Panel(a) of Fig.~\ref{fig:lc}.
A quick look also reveals that flux changes in different bands are apparently concurrent (except for the radio) and a more careful inspection also shows that the degree of change differs between the EM bands. 
We must note that data in different EM bands are extracted using different time-bin durations and thus, are not strictly simultaneous. 
For example, at $\gamma$-ray energies, a 10-day duration is used to ensure sufficient counts for a data point, while such good measurements take just a few minutes at optical and NIR bands. \\
\\

\subsubsection{Segment 1 (MJD 54493 -- MJD 55115)}
\noindent
The first segment includes four flaring states.
Visual inspection of the multi-wavelength light curves shows a close correlation between the $\gamma$-ray, optical, NIR, and radio emissions. 
The optical and NIR bands also brighten during the high activity of the last three $\gamma$-ray high states.
The observed increase in radio flux seems to start quasi-simultaneously with the third $\gamma$-ray high state and reached its peak during the period of the fourth $\gamma$-ray high state.

\subsubsection{Segment 2 (MJD 55122 -- MJD 56202)}
\noindent
The source showed intense activity in this period, with four major $\gamma$-ray flaring states.
The normalized variability amplitudes are not consistent between the bands.
In addition, there are seasonal gaps in optical and NIR light curves during the third of the four $\gamma$-ray flaring states due to the target's proximity to the Sun.
Therefore, it is difficult to tell by inspection whether there are any correlations between different bands. 

\subsubsection{Segment 3 (MJD 56202 -- MJD 56992)}
\noindent
In this segment, the fluctuating trends in the optical and NIR light curves visually coincide with the single $\gamma$-ray flaring state but the radio flux density is not as clearly correlated.

\subsubsection{Segment 4 (MJD 56992 --MJD 58104)}
\noindent
During the first high state of segment 4, four rapid sub-flares were already identified as A (MJD 57100 to MJD 57150), B (MJD 57150 to MJD 57180), C (MJD 57208 to MJD 57235), and D (MJD 57235 to MJD 57260) in~\citet{prince.19.apj}.
The four flares described in their paper appear as four peaks in our $\gamma$-ray light curve constructed using the 10-day bins.
During the high state period, the optical and NIR emissions also exhibit rapid and complex variations. 
Visual inspection shows that the first three $\gamma$-ray radiation peaks appear almost quasi-simultaneously in the optical band, while the first two peaks also appear almost quasi-simultaneously in the NIR band.
Meanwhile, it can be clearly seen that the strong $\gamma$-ray emission occurs during the rising phase of the radio outburst. 
During the second high state of segment 4, there are three near-simultaneous low-frequency counterparts to the $\gamma$-ray flaring state.
The prominent radio flare shows longer rise and decay time scales than $\gamma$-ray and optical bands.

\begin{figure*}
    \includegraphics[width=8.5cm]{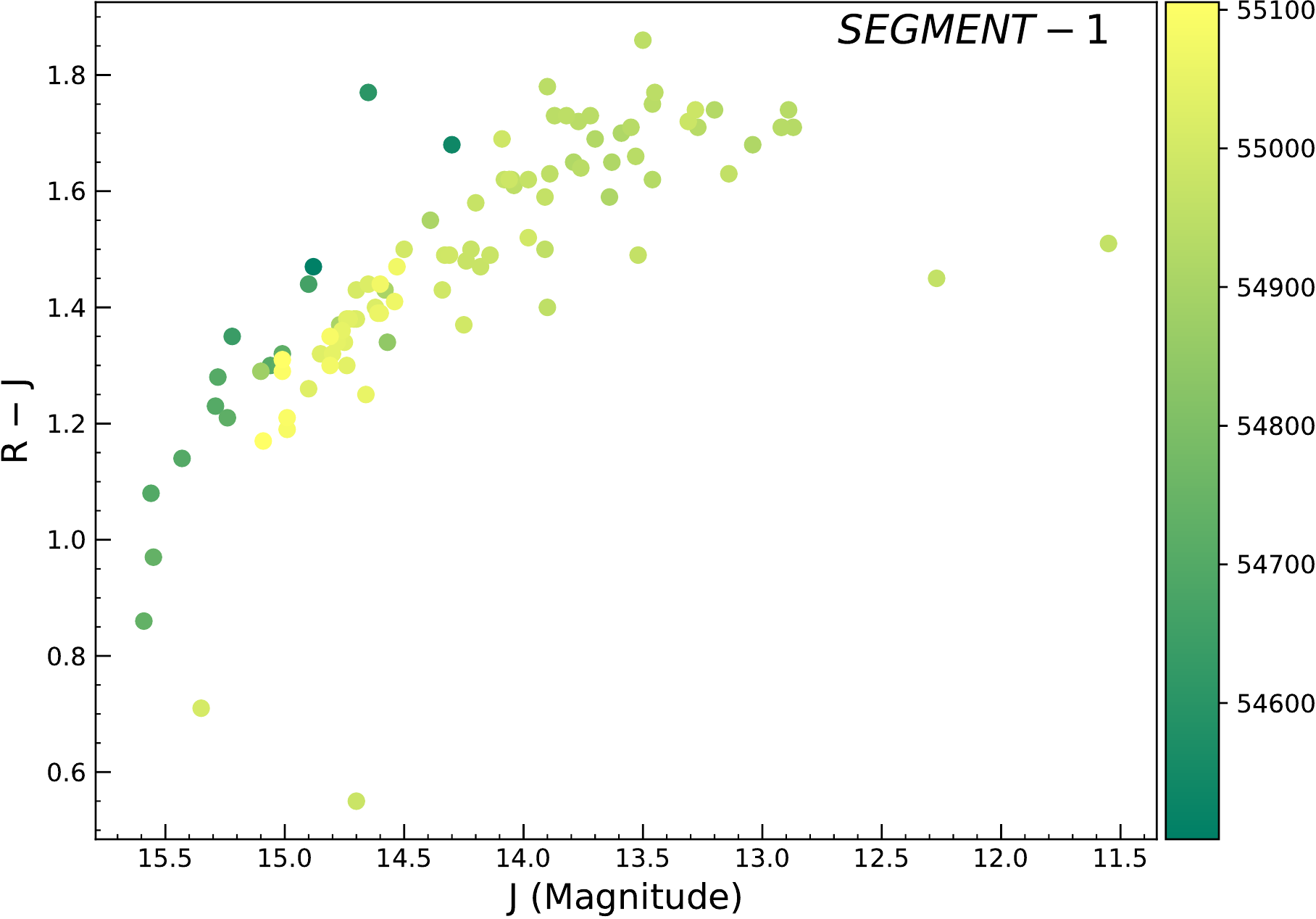}
    \includegraphics[width=8.5cm]{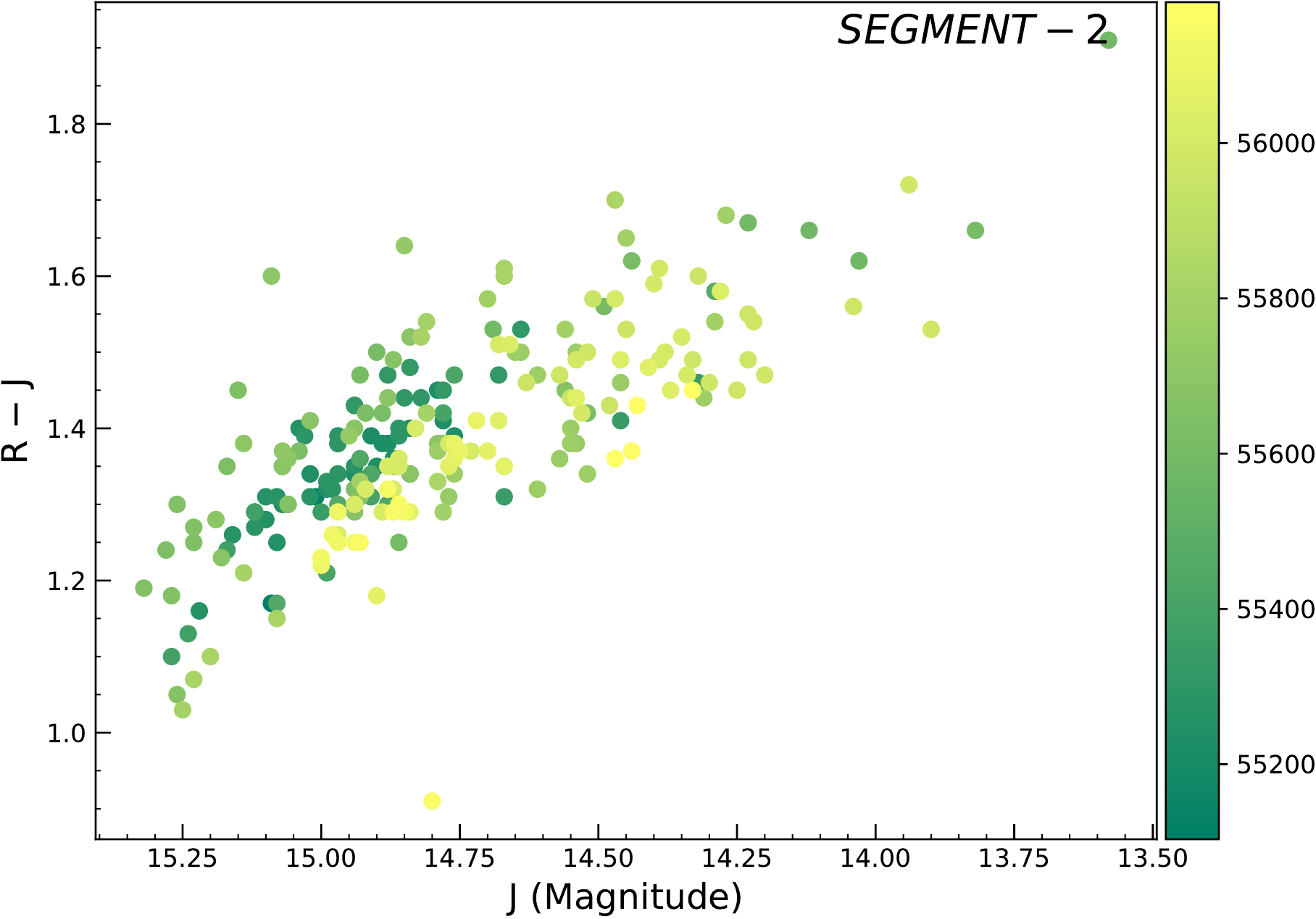}\\
    \includegraphics[width=8.5cm]{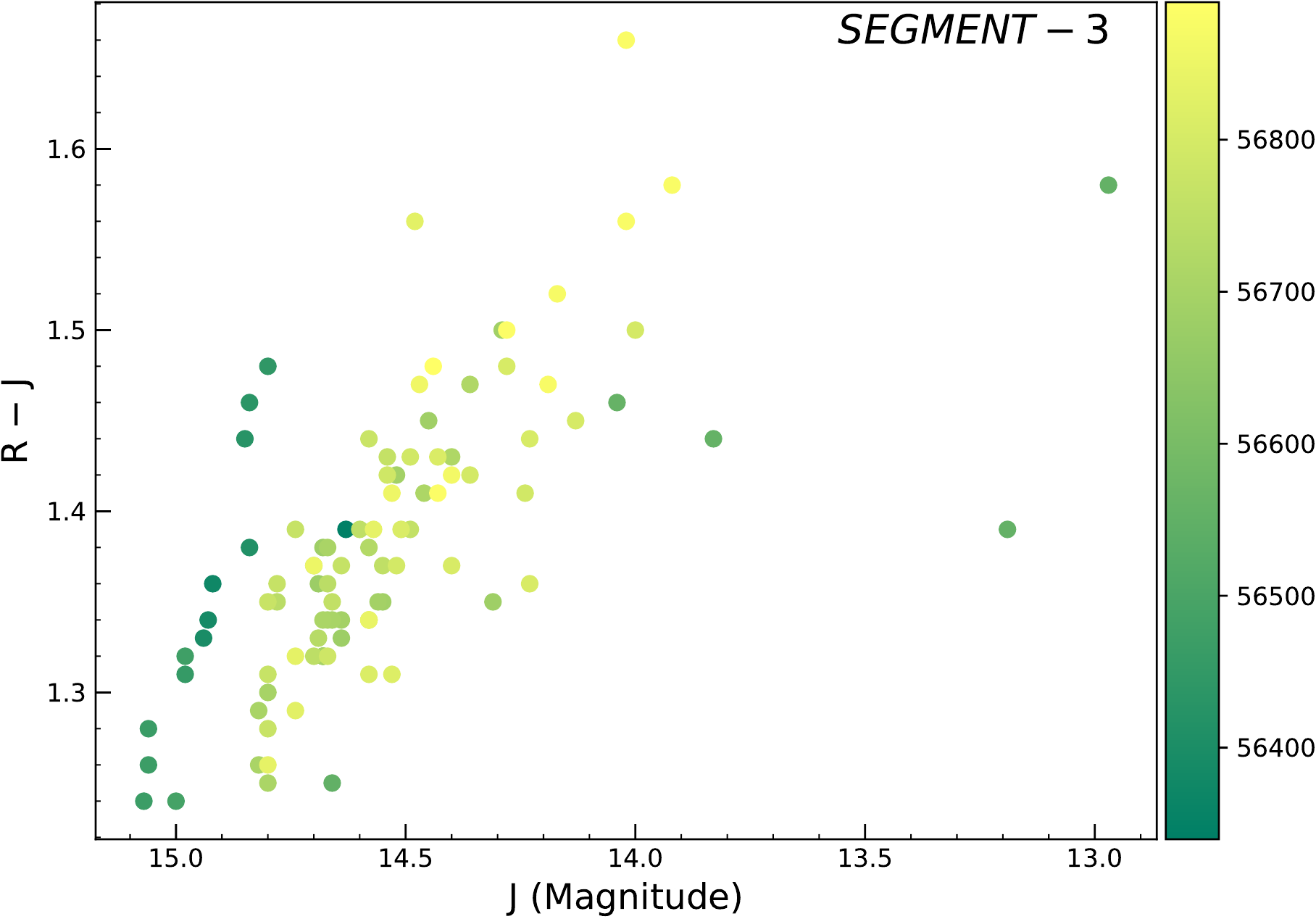}
    \includegraphics[width=8.5cm]{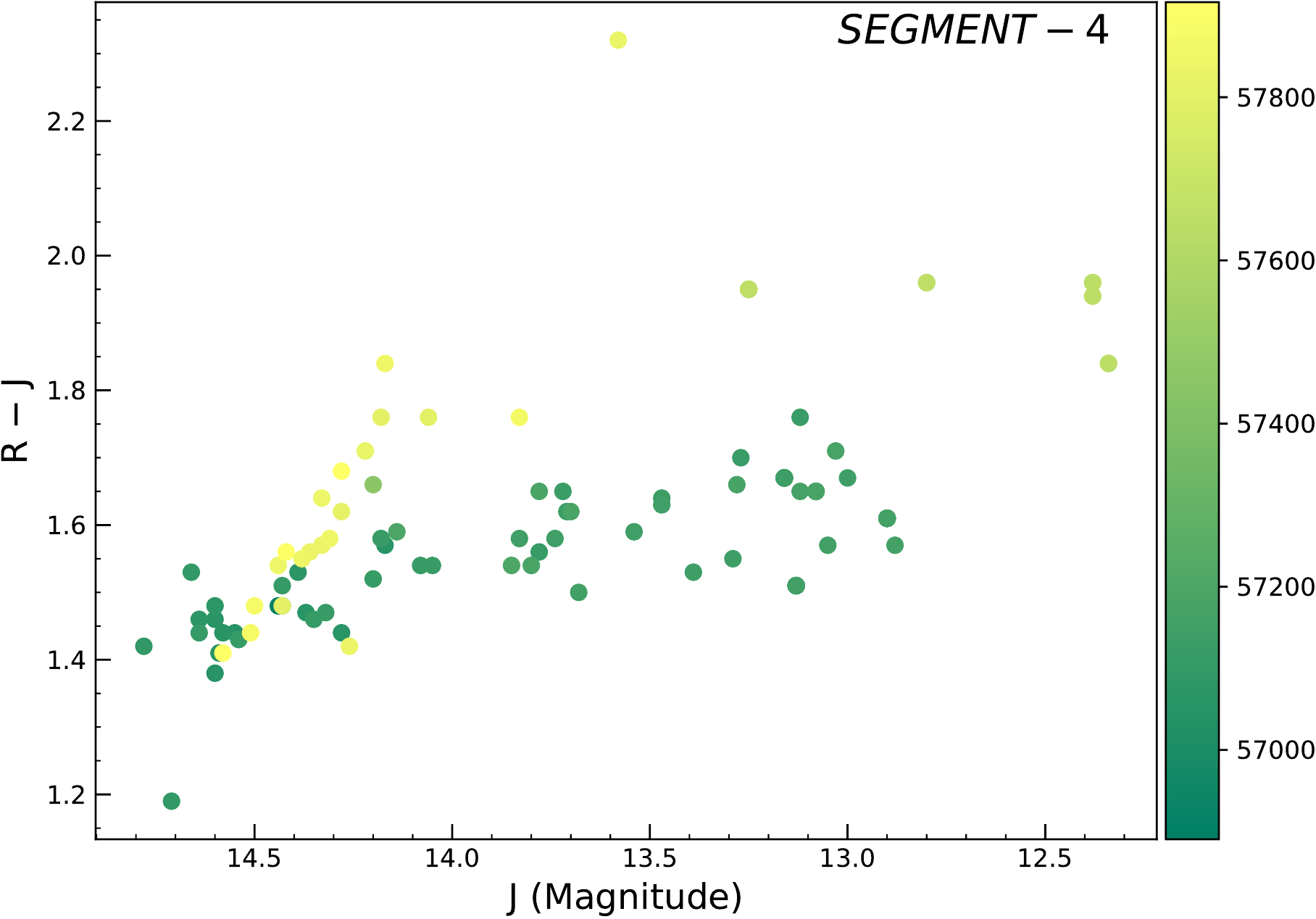}\\
    \caption{Color index (R $-$ J) vs.\ J magnitude in each segment;
    the color bars indicate the progression of time in MJD.}
    \label{fig:color-allsegment}
\end{figure*}

\begin{figure}
    \includegraphics[width=8.5cm]{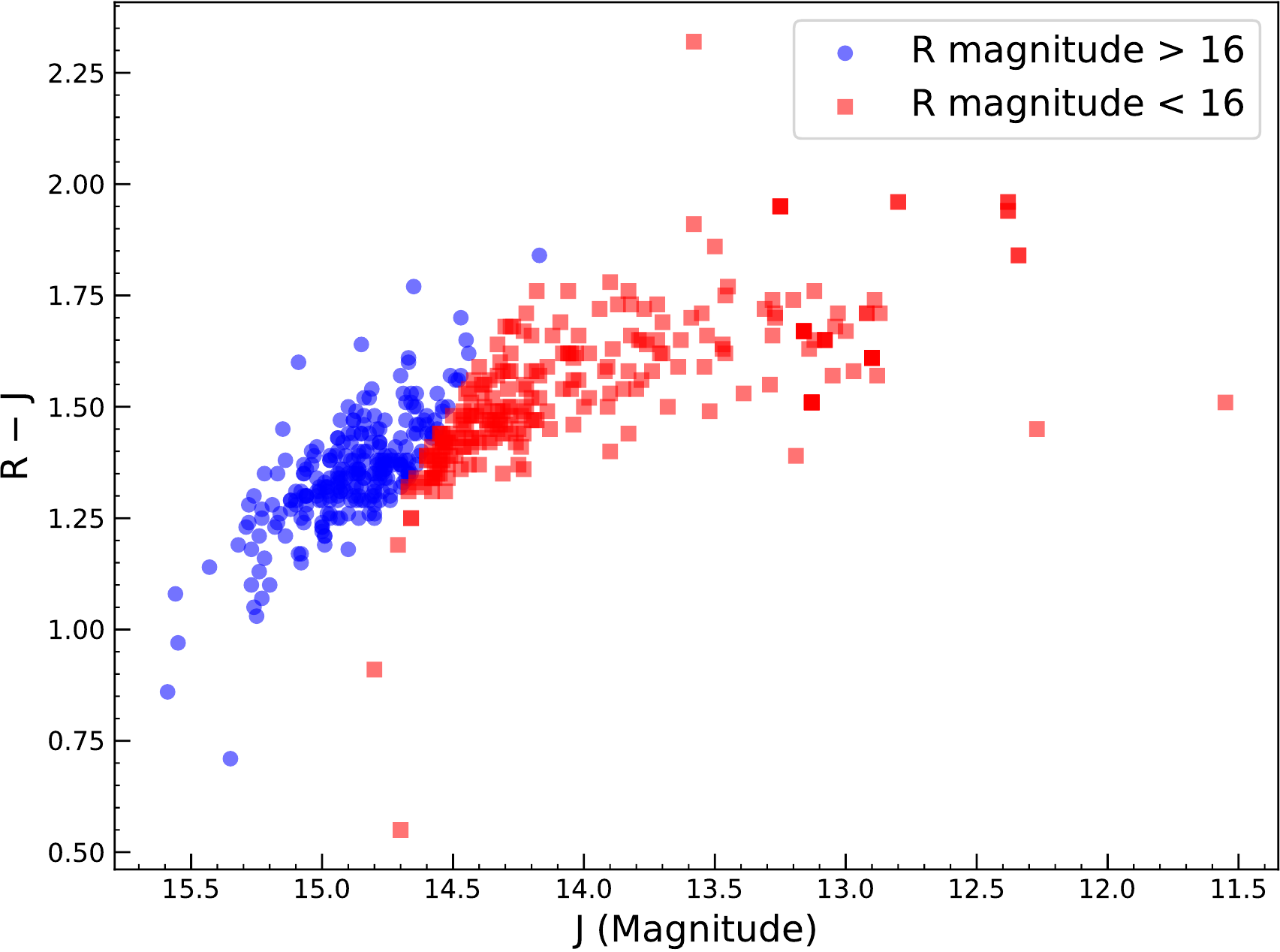}
    \caption{Color index (R $-$ J) vs. J magnitude, grouped by into low flux states (R magnitude $\textgreater$ 16, blue circles) and high flux states (R magnitude $\textless$ 16, red squares). }
    \label{fig:color-low-high}
\end{figure}

\subsection{Spectral Variations}
\noindent
We calculate the (R$-$J) color index of the datasets to examine color variability.
Some studies indicate that the observed optical/NIR color changes depending on the current combination of the relativistic jet's synchrotron emission and the accretion disk's thermal emission~\citep{isler.17.apj,sarkar.19.apj}. 
Compared to the complex color trends found in studies of some other blazars, PKS 1510$-$089 shows a simple color behavior.
It is clear that the color trends of all segments exhibit a redder-when-brighter (RWB) behavior (Fig.~\ref{fig:color-allsegment}), i.e., the color index increases with increasing brightness. \\
\\
We arbitrarily define a low state when the R-band magnitude is greater than 16, and vice versa as the high state.
During the optical low state observation periods, the R-band and J-band light curves do not vary rapidly or drastically.
When the optical emission is in its low state, the accretion disk emission can dominate and the accretion disk luminosity fluctuates over a wide range of time scales~\citep{Lira.11.mn}.
Therefore, we investigate the correlation between the (R$-$J) color index and the J magnitude from another perspective using the above state division, which is displayed in Fig.~\ref{fig:color-low-high}.

\begin{figure}
    \includegraphics[width=8.5cm]{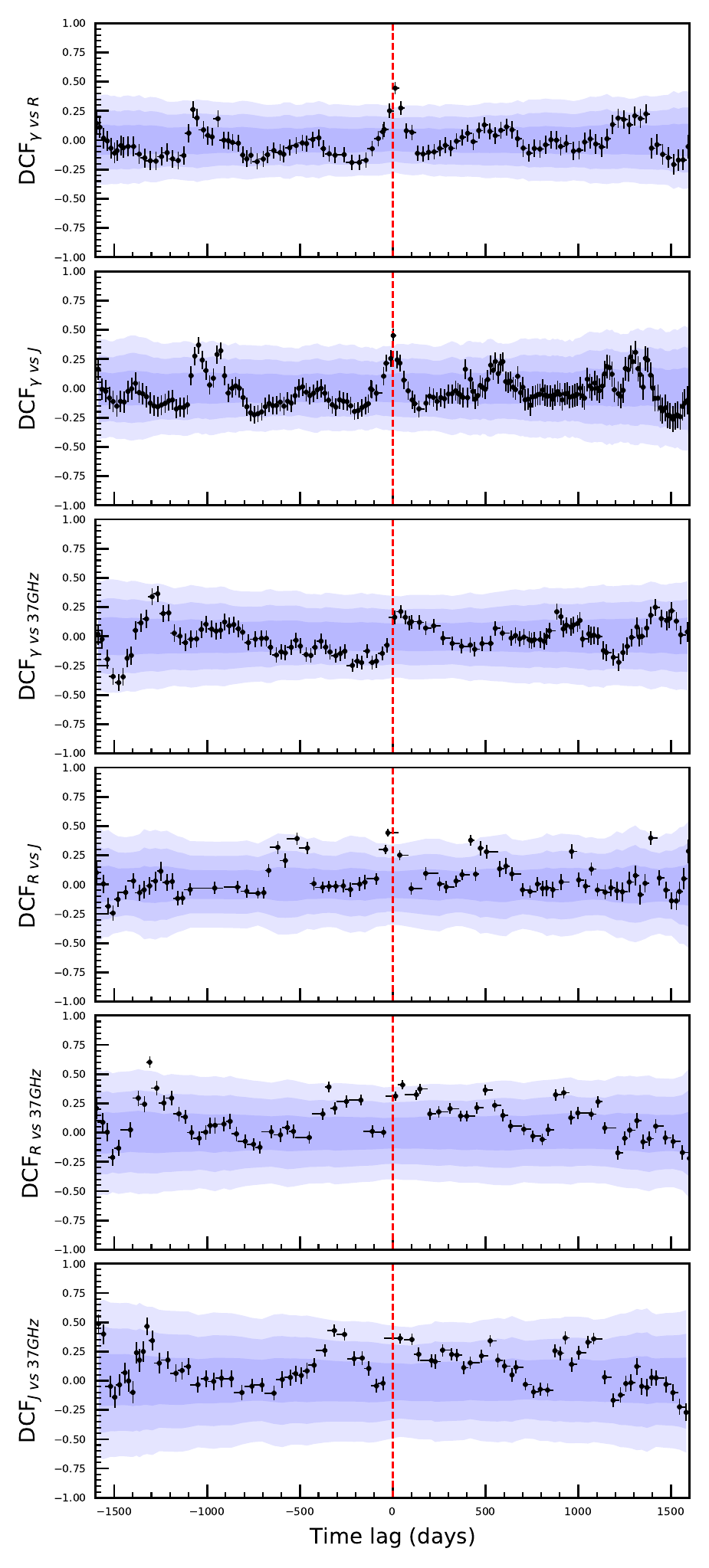}
    \caption{ZDCFs between different wavebands over the entire time of the observations; 
    a ZDCF peak at positive time lag means the LC1 precedes the LC2 where panels are labeled as `LC1' vs `LC2'.
    The color contours denote the distribution of random cross-correlations obtained by Monte Carlo simulations; from dark to light these are 1$\sigma$, 2$\sigma$, and 3$\sigma$.}
    \label{fig:zdcf-merge-result}
\end{figure}

\section{VARIABILITY ANALYSIS AND RESULTS}
\subsection{Method}

\begin{figure*}
    \includegraphics[width=18cm]{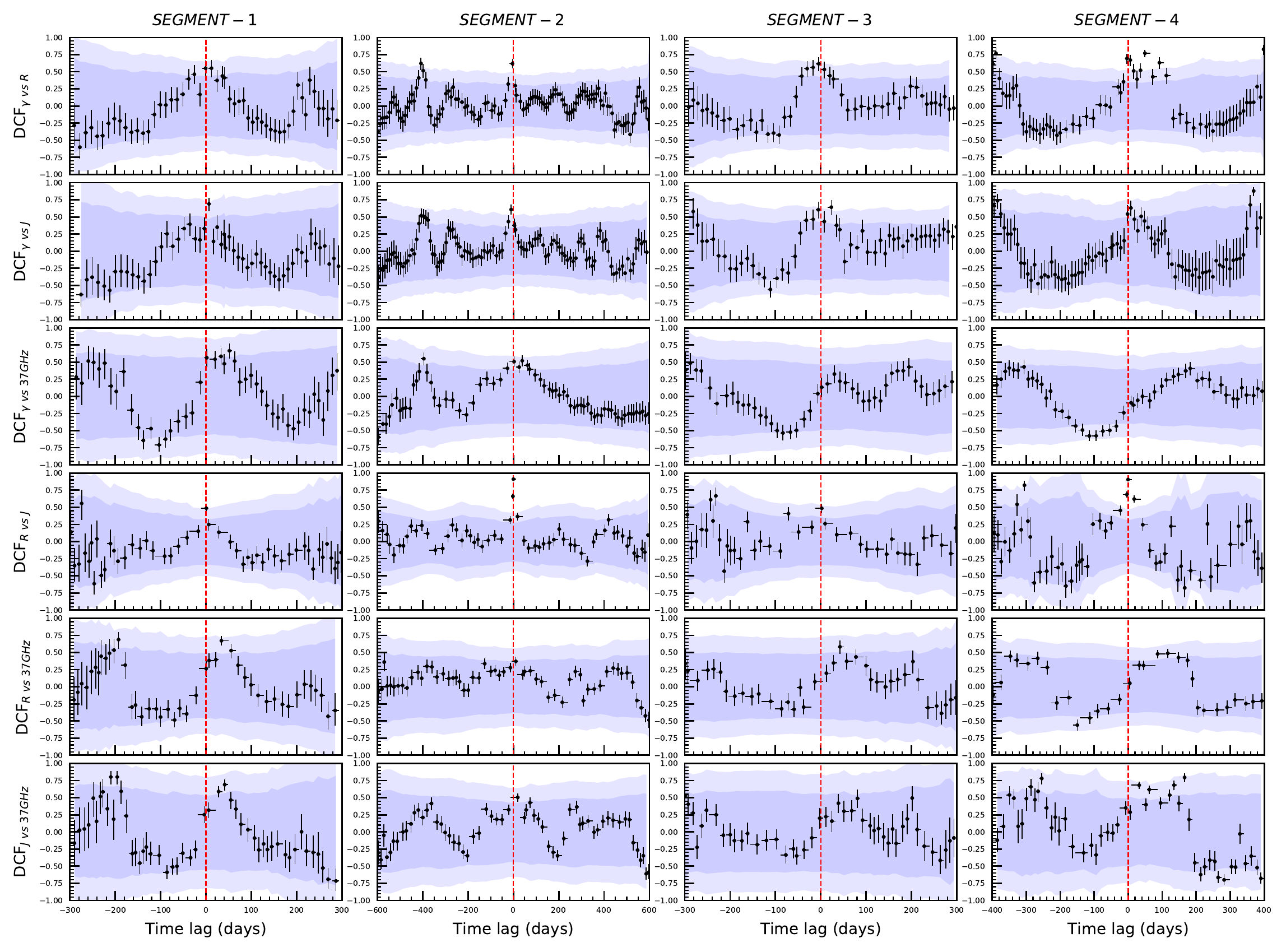}
    \caption{ZDCFs between different wavebands in the four temporal segments; 
    a ZDCF peak at positive time lag means the LC1 precedes the LC2 where panels are labeled as 'LC1' vs 'LC2'.
    The color contours denote the distribution of random cross-correlations obtained by Monte Carlo simulations; here the dark and light are 2$\sigma$ and 3$\sigma$, respectively.}
    \label{fig:zdcf-result}
\end{figure*}

\noindent
Traditional timing analysis requires that a time series is sampled evenly in the time domain.
In astronomical observations, due to the observation schedule, weather and technical issues, 
it is more common that the sampling of the light curve of each campaign is uneven.
We apply the z-transformed discrete correlation function (ZDCF) method~\citep{alexander.97,alexander.13.arxiv} to search for the correlation and time lag between the light curves of each pair of bands.
This method uses equal population binning and Fisher's z-transform~\citep[][and references therein]{kendall1969advanced,kendall1973advanced} to correct several biases of the discrete correlation function (DCF) method~\citep{edelson.88.apj}.
This method also ensures that the statistical significance of each bin is high enough by changing the bin width, a form of adaptive binning.
The specific binning algorithm we used is described in detail in ~\citet{kramarenko.22.mn}.
After binning, the Pearson's correlation coefficient ($r$) is calculated and transformed to the z-space:

\begin{equation}
    z=\frac{1}{2} \log \left(\frac{1+r}{1-r}\right), ~ \zeta=\frac{1}{2} \log \left(\frac{1+\rho}{1-\rho}\right), ~ r=\tanh z,
\label{ztra1}
\end{equation} 
where $\rho$ is the unknown population correlation coefficient of the bin. The ZDCF method uses the Ansatz that
 $\rho = r$ for the transformation. \\
\\
The mean of $z$ ($\bar{z}$) and the variance of $z$ ($s_{z}$) are estimated by 
\begin{equation}
    \bar{z}=\zeta+\frac{\rho}{2(n-1)} \times\left[1+\frac{5+\rho^{2}}{4(n-1)}+\frac{11+2 \rho^{2}+3 \rho^{4}}{8(n-1)^{2}}+\cdots\right],
\label{ztra2}
\end{equation}
and
\begin{equation}
    s_{z}^{2}=\frac{1}{n-1}\left[1+\frac{4-\rho^{2}}{2(n-1)}+\frac{22-6 \rho^{2}-3 \rho^{4}}{6(n-1)^{2}}+\cdots\right].
\label{ztra3}
\end{equation}
The error on the ZDCF values is determined from 300 Monte Carlo simulations where light curves
are generated by randomly adding error to each data point according to its observed error, followed by finding
their ZDCF estimates. The mean and variance are then estimated in the z-space, which is then transformed
back to the real space to estimate the error \citep[see Eqn.\ (8) of][]{alexander.13.arxiv}.\\
\\
To estimate the significance of the correlation coefficients, we follow the Monte Carlo method described by \citet{max-b.14.mn} based on the simulated light curves having variability and statistical properties similar to the observed light curve.
In this work, we simulated a total of 1000 light curves with the same power spectral density (PSD) and probability density function (PDF) as the observed light curves in each band using the algorithm presented by~\citet{emmanoulopoulos.13.mn}, as realized by the DELightcurveSimulation code~\citep{connolly.15.arxiv}\footnote{\url{https://github.com/samconnolly/DELightcurveSimulation}}.
We estimate the underlying PSD by fitting a smoothly bending power-law model plus a constant, c:

\begin{equation}
\mathcal{P}(f ; \bm{\gamma}, c)=\frac{A f^{-\alpha_{\text {low }}}}{1+\left(f / f_{\text {bend }}\right)^{\alpha_{\text {high }}-\alpha_{\text {low }}}}+c,
\end{equation}
where $\bm{\gamma}=\left(A, f_{\text {bend }}, \alpha_{\text {low }}, \alpha_{\text {high }}\right)$ represents the model parameters, which are normalization, bend frequency, low-frequency slope, and high-frequency slope, respectively; $c$ is an additional constant Poisson noise.
The optimal parameters of this PSD model were obtained by maximum likelihood using the Basin-Hopping algorithm and the Nelder-Mead minimization algorithm provided by the Python package SciPy.
As for the PDF model, we use a mixed distribution model consisting of a gamma distribution and a log-normal distribution:
\begin{equation}
f_{\text {mix }}(x)=w_{\Gamma} \frac{\theta^{-\kappa} \mathrm{e}^{-x / \theta} x^{\kappa-1}}{\Gamma(\kappa)}+w_{\ln \mathcal{N}} \frac{\mathrm{e}^{-(\ln x-\mu)^{2} /\left(2 \sigma^{2}\right)}}{\sqrt{2 \pi} x \sigma},
\end{equation}
where $\kappa$ and $\theta$ are the shape and the scale parameters of the gamma distribution;
$\mu$ and $\sigma^{2}$ represent the mean and the variance of the log-normal distribution;
 $w_{\Gamma}$ and $w_{\ln \mathcal{N}}$ are the weights of the two distributions, respectively, where their sum is 1. 
The optimal PDF model parameters are  obtained through a maximum likelihood analysis using the BFGS (Broyden-Fletcher-Goldfarb-Shannon) minimization algorithm in the Python package SciPy.
Table~\ref{tab:para} lists the best-fit parameters for both models across all bands. \\

\begin{table*}[]
    \centering
    \caption{The best-fit parameters for both models across all bands}
\begin{tabular}{l|lllll|llllll}
\hline
 &\multicolumn{5}{l|}{PSD}   &\multicolumn{6}{l}{PDF} \\ \hline
       &A(Hz$^{-1}$)   &$f_{\text {bend }}$(Hz)   &$\alpha_{\text {low }}$   &$\alpha_{\text {high }}$  &c(Hz$^{-1}$)   &$\kappa$   &$\theta$   &$\mu$   &$\sigma$   &$w_{\Gamma}$   & $w_{\ln \mathcal{N}}$ \\ \hline
$\gamma$-ray   &22.8056   &0.0070   &0.1202   &1.7735   &0.1288   &5.9916   &6.01437   &0.7580   &4.5668   &0.0994   &0.9006  \\ \hline
R   &0.0814   &0.1789   &0.8749   &5.8442   &0.0542   &3.8340   &0.7921   &0.2269   &1.3622   &0.4850   &0.515   \\ \hline
J   &0.1263   &0.1048   &0.9620   &2.6107   &0.0293   &3.0122   &1.5823   &0.2814   &2.1324   &0.6625   &0.3375  \\  \hline
37~GHz   &3.4348   &0.0015   &0.3314   &1.8827   &0.0070   &5.3792   &0.6663   &0.1914   &2.741   &1.1674   & $-0.1674$ \\  \hline
\end{tabular}
    \label{tab:para}
\end{table*}

\noindent
We calculated the correlation coefficients between the simulated light curves in the exact same way as done for the observed data.
After that, significance levels for observed data correlation coefficients are derived from the distribution of simulated correlation coefficients for each time lag.
It should be noted that significance estimated this way assumes completely random variations in the different bands which is not true; e.g., NIR J and optical R bands both are dominated by the same synchrotron component. Similarly, $\gamma$-ray fluxes are related to the optical ones if the same particle distribution is responsible for both the synchrotron and inverse Compton components. The synchrotron radio emission probably has contributions from a much bigger region and also encodes opacity effects. Thus, the quoted significances are overestimates in general.

\subsection{Results}
\noindent
The ZDCF results for the entire span of these observations and for the four individual segments are shown in Figs.~\ref{fig:zdcf-merge-result} and \ref{fig:zdcf-result}, respectively. 
The time lags at which the ZDCFs peak are reported in Table ~\ref{tab:lagtime}.
Any ZDCF peak within the sampling time resolution is not considered to indicate an actual time lag between the two bands' emissions. 
We only consider the highest ZDCF peak closest to 0 lag and discard any peaks near the edges of the temporal span considered (e.g., $\pm 300$ d for segment1).

\subsubsection{Interband correlations over the entire time}
\noindent
We first consider interband correlations over the entire span of these observations.
We find correlations close to zero lag with significance larger than 2$\sigma$ between all frequency bands except for $\gamma$-ray and 37 GHz, with four of them exceeding 3$\sigma$: $\gamma$-ray and R-band, $\gamma$-ray and J-band, R-band and J-band, and R-band and 37 GHz~(Fig.~\ref{fig:zdcf-merge-result}).
The result that these ZDCF peaks near zero lag rarely exceed 0.5 can be understood as the cumulative effect of incorporating the gaps in the R and J band data between each segment.
We know that even for the same blazar, there can be significant differences in multi-band correlations during different observational intervals.
Therefore, we study the correlation of light curves between different bands over the four segments described above, thereby hoping to minimize the impact of missing data on the intrinsic correlations. 

\begin{table*}
   \centering
   \caption{Time lags, in days, of the ZDCF peaks between different bands}
   \label{tab:lagtime}
   \begin{tabular}{rrrrrrrr}
      \hline
    Light curves                &  ENTIRE TIME   &  SEGMENT 1         &   SEGMENT 2   &   SEGMENT 3   &   SEGMENT 4\\
      \hline

        $\gamma$-ray versus R        &  $13.27_{-4.40}^{+22.60}$      &  $-0.81_{-6.85}^{+2.57}$  &   $-4.39_{-13.72}^{+2.28}$  &  $-4.58_{-6.52}^{+2.96}$       &  $-3.82_{-2.52}^{+4.47}$     \\
        $\gamma$-ray versus J       &  $2.81_{-3.17}^{+16.52}$     &  $6.57_{-4.23}^{+2.76} $  &   $9.85_{-6.94}^{+2.24} $   &  $-5.42_{-5.01}^{+3.79}$       &  $8.13_{-2.52}^{+4.47}$      \\
        $\gamma$-ray versus 37~GHz   & $--$                         &  $51.65_{-3.65}^{+5.24}$  &   $39.03_{-5.79}^{+3.77}$   &  $--$                          &  $--$                        \\
        R versus J              & $-25.97_{-8.08}^{+55.87}$   &  $0.81_{-11.81}^{+2.17}$  &   $0.16_{-0.16}^{+11.78}$   &  $2.05_{-14.01}^{+3.75}$       &  $-0.42_{-1.53}^{+10.52}$    \\
        R versus 37 GHz         & $53.5_{-28.02}^{+5.96}$     &  $34.01_{-3.48}^{+15.63}$ &   $10.89_{-18.07}^{+3.00}$  &  $42.20_{-2.58}^{+8.25}$       & $--$                         \\
        J versus 37 GHz         & $39.64_{-84.76}^{+16.01}$     &  $41.25_{-2.43}^{+7.56} $ &   $19.18_{-32.35}^{+6.55}$  &  $38.84_{-3.12}^{+4.91}$       & $--$                        \\
     \hline
   \end{tabular}
\end{table*}

\subsubsection{Interband correlations for each segment}

\noindent
We now consider the results for each segment in examining the correlations between the bands.
In most cases, the significance of the highest peak ZDCF peak closest to zero lag is greater than 2$\sigma$ and has a value exceeding 0.6.\\

\noindent
\emph{$\gamma$-ray versus optical/NIR:}\\  
There are significant correlations between the $\gamma$-ray photon flux and the optical R-band and NIR J-band flux densities for all segments. 
Recall that we use a time bin of 10 d to reconstruct the $\gamma$-ray light curve. 
The results given in the first two rows of Fig.\ \ref{fig:zdcf-result} and Table ~\ref{tab:lagtime} show that all these ZDCF peaks are within that 10 day interval and hence we measure essentially no time lag between the $\gamma$-ray and optical/NIR emissions in every segment.\\

\noindent
\emph{$\gamma$-ray versus radio:}\\
The $\gamma$-ray versus radio ZDCF peaks at non-zero time lags in the first two segments, with the $\gamma$-ray emission leading the radio emission by about 50 days in segment 1 and about 40 days in segment 2.
In segments 3 and 4, no statistically robust correlations between the $\gamma$-ray and 37 GHz bands is seen.\\ 

\noindent
\emph{Optical versus NIR:}\\
For all segments, the optical versus NIR ZDCF peaks are very close zero time lags, indicating that the R-band and J-band emissions are simultaneous or at least that any delay between these two light curves occurs on the time-scale smaller than the cadence of observations.\\

\noindent
\emph{Optical/NIR versus radio:}\\
In segment 1 and segment 3, both the optical and NIR versus radio ZDCF peaks correspond to about 40 days, with the optical/NIR emission leading the radio emission.
During the MJD 55800 $\sim$ MJD 55900 portion of segment 2 the radio emission rises rapidly, and optical and NIR data for this period happen to be missing because of the source's proximity to the Sun. 
Therefore it is not surprising that the ZDCF peak for segment 2 shows only a weak correlation coefficient between these bands and the errors of the time lags corresponding to those peaks are very large.
In segment 4, as mentioned above, the radio flares show longer rise and decay timescales than do the changes in the $\gamma$-ray and optical/NIR bands. 
Therefore for this fourth temporal segment, the optical and NIR correlations and broad nominal positive lags we obtain to the radio emission by applying mathematical methods such as ZDCF are unconvincing.

\section{Quasi-periodic Oscillation Analysis} 
\noindent
 Over the last 15 years or so a modest number of QPO detections have been reported in blazars in different EM bands. 
 \citep[e.g.,][and references therein]{2014JApA...35..307G,Gupta:2017ukw,sarkar.19.apj,2022MNRAS.510.3641R,2022MNRAS.513.5238R}. 
 PKS 1510$-$089 is among a few blazars which has had published claims of detection of QPOs in different EM bands in different temporal spans.
 In 22 and 37 GHz radio bands data taken during 1995 -- 2005, gave indications of QPOs with two periods, P1 = 0.92 yr and P2 = 1.84 yrs, where P1 is a harmonic of P2 \citep{2008AJ....135.2212X}. 
 In UMRAO (University of Michigan Radio Astronomical Observatory) data taken in 8 and 14.5 GHz during 1974 -- 2011, indications of multiple QPO periods in the period range 430 to 1080 days were found \citep{2021JApA...42...92L}. 
 In a poorly sampled optical data taken during 1999 -- 2001, QPO with period of $\sim$ 0.92 yr was suggested \citep{2002MNRAS.334..459X}, 
 though the data train was not long enough to make this a convincing case. 
 Other poorly sampled NIR K band and optical R band light curves of the source taken during the longer span of 2006 -- 2014 apparently displayed multiple periods ranging from 203 to 490 days \citep{2016AJ....151...54S}. 
 In the 2006 -- 2014 Fermi-LAT light curve of this blazar, a QPO with a period of $\sim$ 115 days was claimed \citep{2016AJ....151...54S,castignani.17.aa}.
 Recently, transient $\gamma-$ray QPOs with periods of 3.6 days and 92 days in Fermi-LAT data of the source were claimed in two different portions of the light curve \citep{2022MNRAS.510.3641R}.\\
\\
In light of these previous QPO signals from the blazar PKS 1510$-$089 we now consider for further analysis the independent new data we have presented here: the Mets{\"a}hovi  37 GHz radio light curve between 2008 -- 2018  (plotted in the bottom panel of Fig.\ \ref{fig:lc}). 
Modulations indicating a possible quasi-periodic component might be identified by visually inspecting 
the flux variations with time. We used several methods, the Generalized Lomb-Scargle periodogram (GLSP),
REDFIT, and the Weighted Wavelet Z-transform (WWZ) to assess the possibility of quasi-periodicity in  the frequency domain (and also, for WWZ, the time domain). 
All of these methods can be used for unevenly sampled data which is  normally the case for  
radio light curves observed over the range of few years to decades. \\ 

\begin{figure}
    \centering
    \includegraphics[scale=0.5]{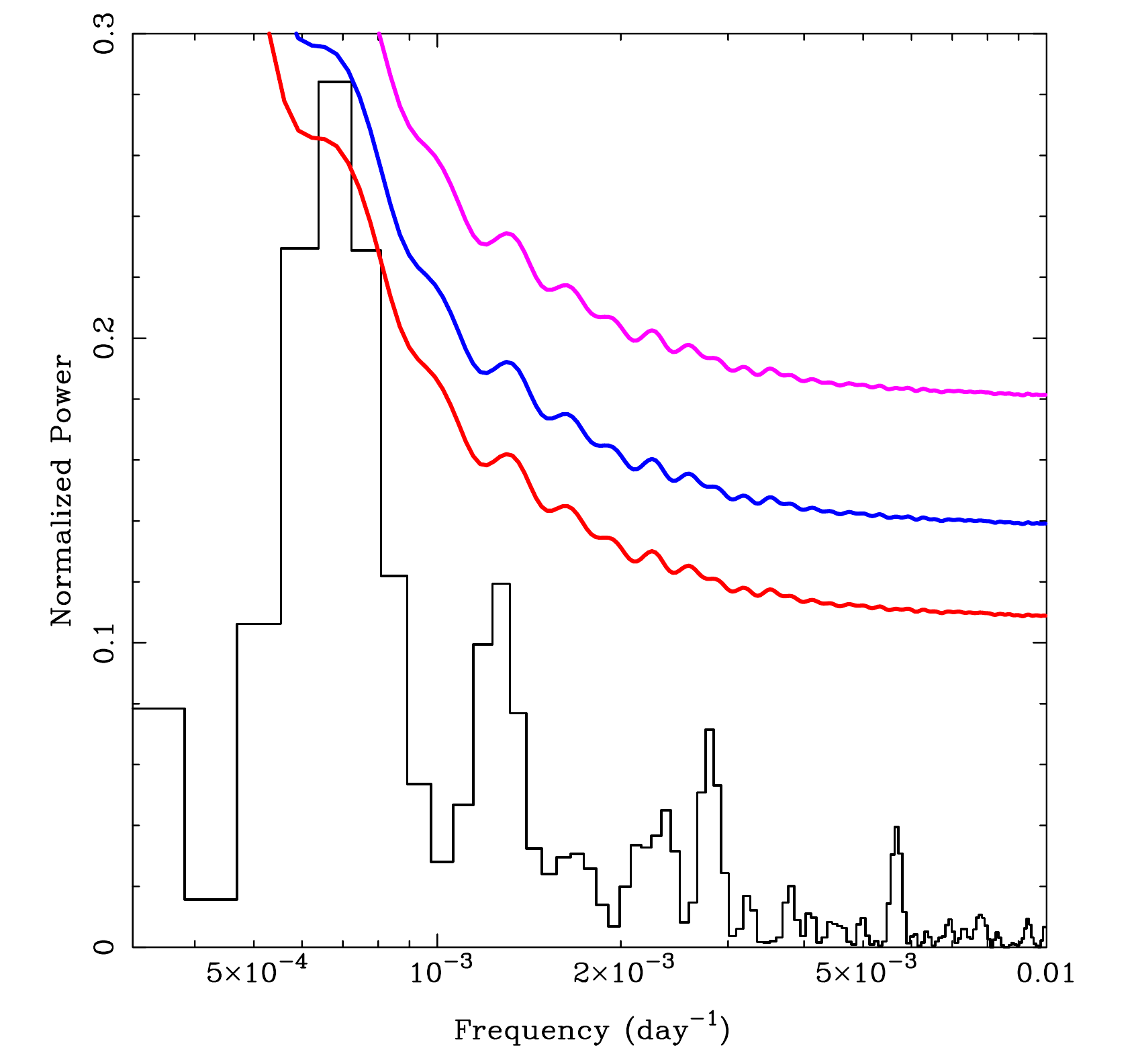}
    \caption{GLSP of the radio light curve of PKS 1510$-$089 for the duration {\bf 2008--2018.} The black histogram denotes the normalized GLSP. The solid red, blue and magenta curves
    denote 3$\sigma$, 4$\sigma$ and 5$\sigma$ confidence intervals, respectively.}
    \label{fig:glsp}
\end{figure}

\begin{figure*}
    \centering
    \includegraphics[scale=0.5]{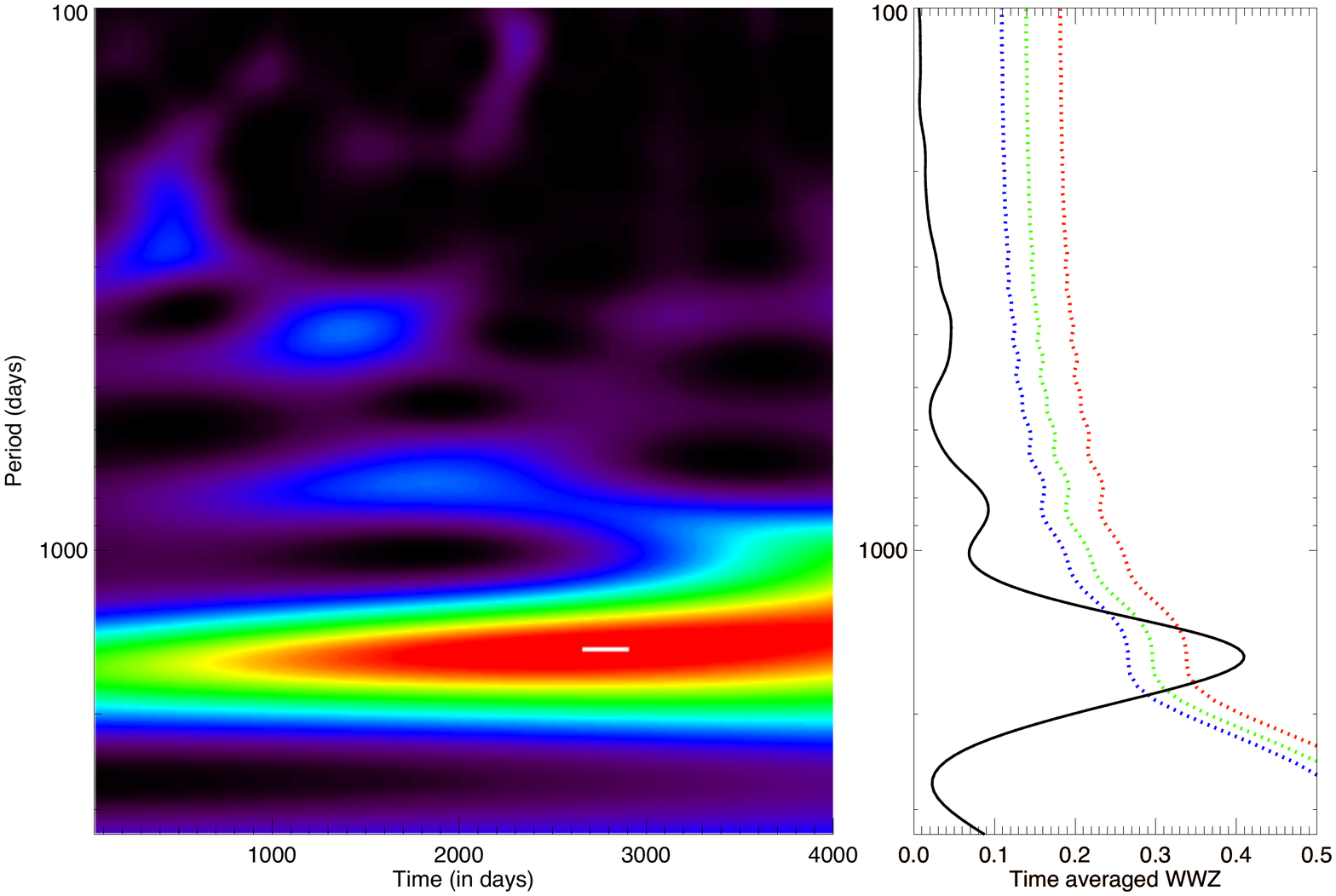}
    \caption{Wavelet result for the {\bf duration} 2008 -- 2018 light curve. Left panel: WWZ plot with
    red denoting the maximum wavelet power which decreases towards violet and black. Right panel: Time averaged WWZ (solid black curve); the blue, green and red dotted curves represent 3$\sigma$,
    4$\sigma$ and 5$\sigma$ confidence intervals, respectively.}
    \label{fig:wwz}
\end{figure*}

\begin{figure}
    \centering
    \includegraphics[scale=0.5]{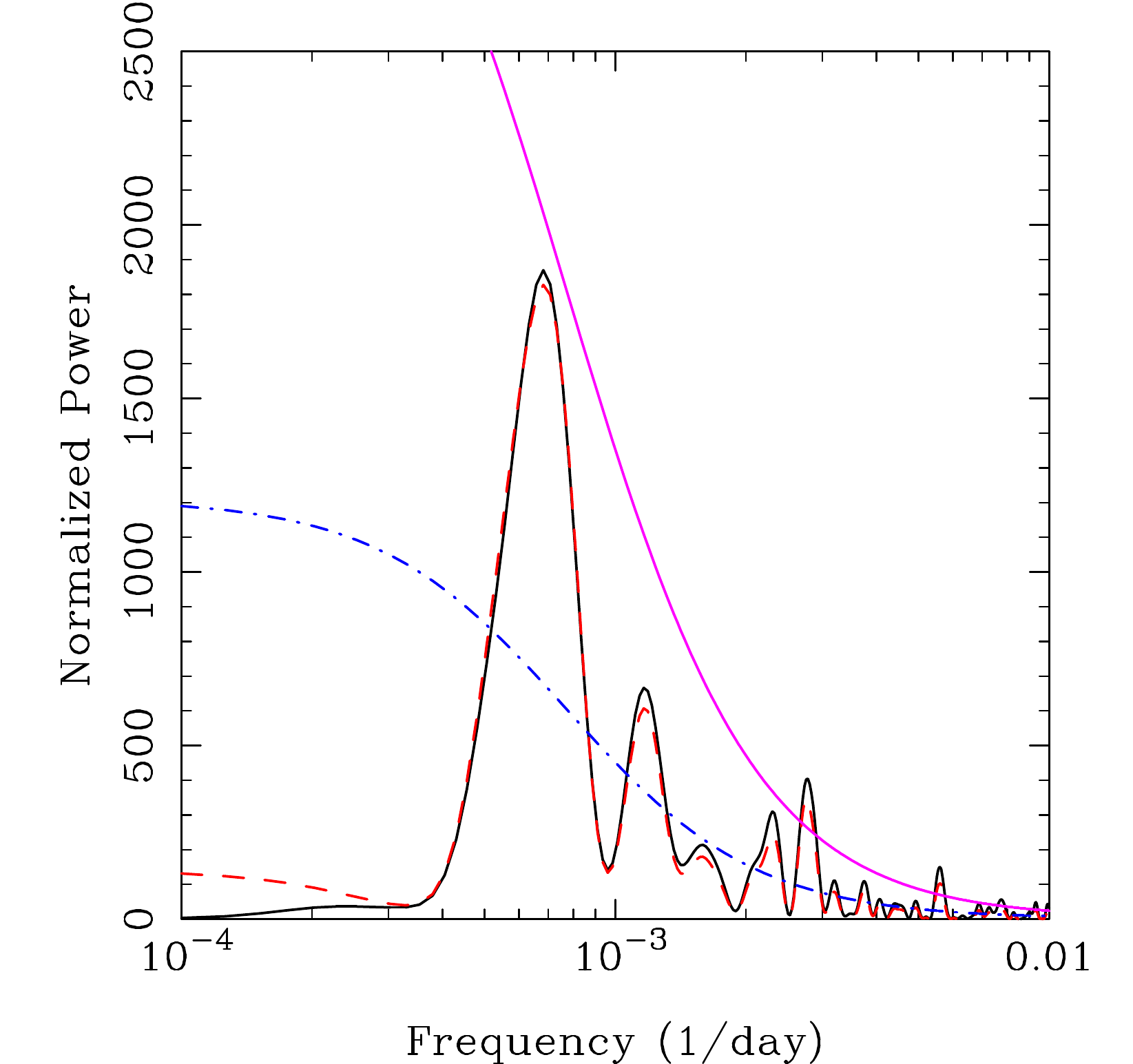}
    \caption{REDFIT result for the 2008 -- 2018 light curve. The black solid curve and the red dashed curve represents the spectrum
    of input data and bias corrected spectrum respectively. The blue dashed-dotted curve denotes the AR(1) theoretical
    spectrum and the solid magenta curve is the 95$\%$ confidence curve.} 
    \label{fig:redfit}
\end{figure}

\noindent 
In this work, we employed the GLSP routine of the {\sc PYASTRONOMY}\footnote{\url{https://github.com/sczesla/PyAstronomy}}
package which is based\ on~\citet{zechmeister.09.aa}. 
Fig.~\ref{fig:glsp} shows the GLSP plotted against frequency (black curve). The red, blue and magenta curves respectively represent 3$\sigma$, 4$\sigma$ 
and 5$\sigma$ significance levels. The peak of the periodogram is found at the frequency of $1566 ^{+439}_{-184}$ days at about 
4$\sigma$ nominal significance.  But we note that with a data train that does not exceed 4000 d, this possible period amounts to less than 3 cycles and so can only be considered to hint at a QPO. \\
\\
The significance of any peak that could be a QPO is estimated by simulating light curves having properties similar to those of the original light curve. To quantify the significance of a peak it must be assumed that the light curve of blazars can be well represented by the stochastic processes occurring in the accretion disk and the associated jets. 
For blazars, the PSD over a range of frequencies {\bf also} can usually be well fit by a power law with negative spectral index $\alpha$ \citep[][and references therein]{Gupta:2017ukw, bhatta:2019bjf}.
So we use a simple power law as the underlying model for the power spectrum 
Any QPO is likely to arise due to coherent disk or jet processes and is different from the stochastic process and thus would appear as
peaks in the power spectrum~\citep{ackermann.12.apj}, the significance of which can be assessed. 
The degree to which any QPO peaks stand above the power law is used to assess the significance of the periodicities \citep[see for details,][]{2021MNRAS.501.5997T}. 
Wavelet functions are commonly used to examine any evolution in frequency and amplitude of a QPO signal~\citep{torrence1998practical}. 
As the data employed in this work are unevenly sampled, we use an improved version of the wavelet approach, namely the WWZ.\\
\\
We calculated the WWZ\footnote{\url{https://www.aavso.org/software-directory}} and time-averaged WWZ for the simulated light curves as well as for the actual data, following \citet{2021MNRAS.501.5997T} to calculate the significance levels. 
If the WWZ amplitude is marginalized over the whole length of the observations, one gets the WWZ as a function of frequency which is essentially the PSD of the WWZ, which should follow the power law and will be distributed as $\chi^2$ with 2 degrees of freedom. 
Again we use a simple power law as the underlying model to fit the time-averaged WWZ.
The left panel of Fig.\ \ref{fig:wwz} shows the wavelet density plot. 
Interestingly, there is only a single peak at $1538^{+24}_{-23}$ days and it is persistent throughout the data stream. 
The right panel of the Fig.~\ref{fig:wwz} plots the time-averaged WWZ against the period.  
The significance of the highly
persistent signal found in the wavelet density plot nominally exceeds 5$\sigma$.\\
\\  
We also employed the REDFIT method\footnote{The code is found at
\url{https://www.marum.de/Prof.-Dr.-michael-schulz/Michael-Schulz-Software.html}}
which essentially compares the data to a first-order auto-regressive
process (AR1)~\citep[][and references therein]{schulz2002redfit,hong.18.aj,gupta.18.aa} and calculates significance based on $\chi^2$ distributions.
In AR analyzes, the flux at a given point of time is compared to the data at the past times using a regressive relation and can include any number of such past values. In the simplest AR1 method, only one value previous to the data is used to compute the theoretical light curve and subsequently, the  AR1 spectrum. For time $t_i$ (i=1,2,...$N$), the AR1 process $r$ can be written as

\begin{eqnarray}
    r (t_i) = \rho_i  r(t_{i-1}) + \epsilon(t_i)\\
    \rho_i = exp(- (t_i - t_{i-1})/\tau),
\end{eqnarray}\\
where $\epsilon$ is the Gaussian noise such that the AR1 process has unit variance and zero mean, $\rho$ denotes the autocorrelation coefficient, and $\tau$ represents the characteristic time scale for the AR1 process. The power spectrum $S_{rr}(f_i)$, corresponding to the AR1 process described in Eq.\ 4 for frequency $f_i$ up to the Nyquist frequency, $f_{nyq}$,  is given as

\begin{equation}
    S_{rr}(f_i) = S_0 \frac{1-\rho^2}{1-2\rho ~cos(\pi f_i/f_{Nyq}) -\rho^2}
\end{equation}
The estimated values of $\rho$ and $\tau$ for this light curve are found to be 0.92 and 206 days, respectively.
Fig.~\ref{fig:redfit} plots the normalized 
REDFIT results against frequency. The solid black and red dashed curve respectively represent the spectrum of the input data and the bias corrected spectrum. 
  The blue dashed-dotted curve shows the 
theoretical AR(1) spectrum. The magenta curve represents a 2$\sigma$ confidence curve. The peak around 360 days has significance exceeding 
2$\sigma$ but it is suspiciously close to 1 year. The otherwise strongest signal is at $1465^{+56}_{-53}$ days and while it is less than 2$\sigma$ it is consistent 
with the results from GLSP and WWZ methods.

\section{DISCUSSION}

\subsection{Unveiling the physics behind the spectral variations}
\noindent
Recently,~\citet{otro.22.mn} used a powerful statistical tool -- non-negative matrix factorization~\citep[NMF, see][]{paatero1994positive,2014sdmm.book.....I} to reproduce the spectral variability for some blazars over ten years using two to four components.
The photometric and spectropolarimetric data used in their study (taken from Steward Observatory) are partly homologous to the optical R-band data in this paper, with temporal coverages slightly (by less than a year) longer than ours.
This study is critical to understanding the origin of blazar variability.
In PKS 1510$-$089, the spectral variability is explained as the sum of a bright BLR component, a power-law (PL) component accounting for the non-thermal synchrotron radiation of the relativistic jet, and the blue accretion disk contribution that shows slight variation and lower brightness than the BLR and the PL for the jet. 
\citet{otro.22.mn} concluded that the contribution of the jet component is related to color and showed an RWB trend consistent with our results.
For PKS 1510$-$089, which is an LSP FSRQ~\citep{ajello.20.apj} and which has an accretion disk temperature estimated as 40,000 K~\citep{abdo.10.apj}, the synchrotron peak and the thermal accretion disk spectrum peak are located at the NIR and the far UV, respectively.
We note that the thermal disk spectrum is bluer, especially in this source, where the accretion disk temperature is so high.
However, non-thermal photons have a redder spectral shape.
So observed colors are influenced by changes in accretion disk emission, jet emission, or both.
It is reasonable to expect that we would observe a bluer-when-brighter (BWB) trend in the low flux regime if thermal radiation from optical/NIR wavelengths is much stronger than or even swamps the jet radiation.
However, our investigation finds the opposite RWB trend.
We can infer that the Doppler-boosted jet emission dominates the entire radiation process, even the optical band. 
Moreover, in the optical/NIR high flux state, RWB behavior is also displayed throughout the observation period. So we also can infer that the high flux states of optical/NIR emission come from enhanced jet activity,
and increased non-thermal emission results in a reddening of the color.

\subsection{Multi-band cross-correlation analysis}
\noindent
Blazars are the most powerful, persistent extra-galactic broadband sources.
The region responsible for the broadband emission is highly compact -- beyond the resolution limit of modern facilities even for the nearest sources, e.g., M87 and Sgr A*~\citep{eht2.19.apjl,eht4.19.apjl,sageht.22.apjl}. 
However, the jet region is optically thin at NIR and higher energies and thus, temporal flux variability allows us to access the spatial extension buried under the resolution limit and is currently the only way to infer these spatial scales. 
Further, though rapid and strong flux variability is one of the defining characteristics of blazars, these high states are rarely accompanied by the strong spectral changes that would indicate a departure from its SED class.

\subsubsection{$\gamma$-ray versus optical/NIR}
\noindent
The notable correlation between $\gamma$-ray and optical/NIR emissions indicate simultaneous variations (within the sampling intervals) and thus strongly supports the leptonic model.
\citet{ramakrishnan.16.mn} studied 15 blazars and reported two sources with a strong correlation in optical and $\gamma$-ray emissions with zero time delay, one of which is PKS~1510$-$089.
We validated these previous results using high-quality data over a more extended period.
In the leptonic scenario, the high-energy emission is produced by relativistic electrons scattering seed photons from low-energy to high-energy through the inverse Compton (IC) process.
Studies employing SED modeling of PKS 1510$-$089 shows that synchrotron radiation from relativistic electrons can adequately describe the mm to optical emission, while high-energy emissions (X-ray to $\gamma$-ray) are not. 
These previous studies have shown that X-ray and $\gamma$-ray emissions are also an EC process of the BLR and DT photons~\citep{kataoka.08.apj,abdo.10.apj,brown.13.mn,barnacka.14.aap,prince.19.apj}.
Moreover,  ~\citet{prince.19.apj} inferred that the $\gamma$-ray and optical emissions are produced in the same region from the flux doubling times of the $\gamma$-ray and optical and UV light curves.

\begin{figure*}
    \includegraphics[width=17.5cm]{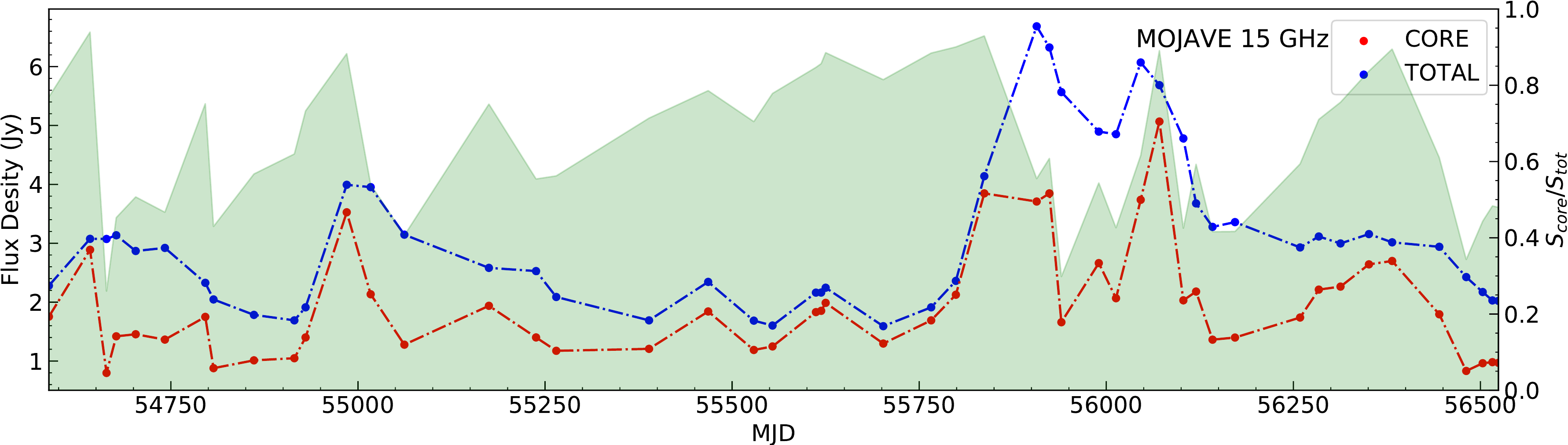}
    \caption{Archived 15 GHz VLBI data light curves for PKS 1510$-$089.
    Red circles represent the fitted core features flux density from~\citet{angelakis.19.aa},
    and blue circles represent the VLBI total flux density from the MOJAVE web page
     {\url{https://www.cv.nrao.edu/MOJAVE/sourcepages/1510-089.shtml}} 
    \citep{lister.18.apjs}.
    The values that make up the upper border of the green shade represent the ratio of the core flux density 
    to the total flux density for each epoch.
    }
    \label{fig:mojave-core-total}
\end{figure*}

\subsubsection{Optical versus NIR}
\noindent
As stated above, the synchrotron radiation of the relativistic jet is the primary source of its optical and NIR radiation~\citep[e.g.][]{blandford.78.book}.
However, additional contributions are expected from thermal radiation from the accretion disk and torus.
A statistical study suggests that the host galaxy and BLR may also contribute significantly to the optical band~\citep{otro.22.mn}.
Quantifying all possible optical and NIR emission contributions has been a challenge.
The combination of the color trend and correlation analysis of optical and NIR light curves is an effective way to understand the radiation mechanism of this source.
We determined that the jet emission dominates the optical and NIR color variability behavior throughout the observation period.\\
\\
The correlation between the optical B-band and NIR J-band observation data of the SMARTS monitoring program from 2008 to 2010 was studied at an early stage~\citep{bonning.12.apj}.
We use the R-band data released by this program and supplementary data from the Steward Observatory to revisit the correlation between the optical and NIR emissions with longer observation times and denser sampling intervals. 
We confirm the results of ~\citet{bonning.12.apj} that there is a very good correlation between the light curves of optical and NIR bands.
Significant correlations with zero time lag are seen between the optical R-band and NIR J-band light curves, indicating that the bulk of these emissions are simultaneous or that any time lag occurs on a timescale smaller than the observation cadence.
It can further be inferred that the main emission region of the optical and NIR photons is co-spatial and originates in the jet.

\subsubsection{$\gamma$-ray versus radio}
\noindent
Several studies have reported the time delay between the $\gamma$-ray emission and the radio emission measured by both a single radio dish and VLBI.
For single dish data, $\gamma$-ray flares occur earlier than radio flares~\citep{max.14.mn,fuhrmann.14.mn}.
And for VLBI data, \citet{pushkare.10.apjl} used a large sample of 183 bright Fermi-detected sources and found that the correlations were strong between the $\gamma$-ray emission and the VLBA core component emission with $\gamma$-ray emission preceding radio emission.
Recently, they again performed the correlation analysis for the larger samples that have been accumulated and found the time delay is in good agreement with their previous results~\citep{kramarenko.22.mn}.
The light curve obtained from single-dish observations or VLBA observations yielded similar results when correlated with the $\gamma$-ray light curve.
The above results are supported for two reasons. 
Firstly, the radio core dominates the radio emission, which makes the core flux density and total flux density light curves' characteristics similar in VLBA observations.
Secondly, the parsec-scale emission region dominates the radio flux density variation, which reflects that the characteristics of total light curves obtained by the single-dish observation and VLBA observation are similar.
Similar characteristics refer to the flares occurring almost quasi-simultaneously, with roughly the same upward and downward trends, although the above characteristics vary from source to source.\\
\\
We can compare the magnitudes of core and total flux densities of PKS 1510$-$089 at multiple epochs when 15 GHz VLBI observations were made by collecting data released by the MOJAVE program~(see Fig.~\ref{fig:mojave-core-total}).
Here we also calculate the values ($S_{\mathrm{core}}/S_{\mathrm{VLBA}}$) that characterize the core dominance degree of each epoch~(also see Fig.~\ref{fig:mojave-core-total}), where $S_{\mathrm{core}}$ and $S_{\mathrm{VLBA}}$ are the core and total flux density, respectively.
It can be clearly seen that this source is core dominated.
In addition, for PKS 1510$-$089, it has been demonstrated that the parsec-scale emission region dominates the radio flux density variability~\citep{orienti.13.mn}.
So, the above $S_{\mathrm{core}}$ and $S_{\mathrm{VLBA}}$ variations are not much different from the radio emission measured by a single antenna.
We therefore can reasonably predict that if there were even more densely sampled VLBI data available we would obtain $\gamma$-ray and radio radiation correlations that are consistent with single-dish data analyses.
We should remember that intensive monitoring is essential to understand the correlation between the cross-band activity for the long-term trends, so our choice of the well-sampled 37 GHz single-dish data to perform the correlation analysis is sensible. \\
\\
In the classical and commonly used shock-in-jet model~\citep[e.g.,][]{marscher.85.apj,turler.2000.aa,fromm.11.aa}, the shock propagates down a conical jet and accelerates the relativistic particles at the shock front.
These particles propagate behind the shock front and lose their energy through different energy loss mechanisms, such as adiabatic expansion and synchrotron radiation. 
The most widely-accepted view is that the moving shock is an intrinsic factor responsible for the observed radio flux density variation.\\
\\
As mentioned above, the current investigations found that $\gamma$-ray emission precedes the radio emission.
Several studies have attempted to interpret this result, suggesting that $\gamma$-ray luminosity variations may be connected to the same shocked radio features and connect these two frequency emissions through the shock~\citep{fuhrmann.14.mn,ramakrishnan.15.mn}. 
Suppose radio and $\gamma$-ray emissions are triggered by shocks propagating in the relativistic jet.
In this scenario, the time lag is related to the distance between the radio core and the location where the $\gamma$-ray emission is produced.
This can also be understood as the unresolved core centroid moving towards the position of the black hole as observing frequency increases~\citep{blandford.79.apj,konigl.81.apj}, i.e., the so-called `core shift' effect.
The $\gamma$-ray photons escape from the jet immediately when the shock arrives, and it takes a while for the shock (perturbation) to propagate farther along the jet until it reaches the $\tau=1$ surface from which radio photons at the specified radio frequency can escape.\\
\\
In our study, we also found that the $\gamma$-ray emission leading the radio emission in segment 1 and segment 2.
Moreover, as shown above, the flux density variability of PKS 1510$-$089 is dominated by the flux density of the core fitted in the VLBI images.
Hence, a possible explanation for these results may be the above scenario involving a shock.
We estimate the separation of the $\gamma$-ray and the 37 GHz emission region following the method and assumption of~\citet{pushkare.10.apjl},
\begin{equation}
    \Delta r=r_{\gamma} - r_{\mathrm{Radio}}=\frac{\delta \Gamma \beta c \Delta t_{\mathrm{Radio}-\gamma}^{\mathrm{obs}}}{1+z}=\frac{\beta_{\mathrm{app}} c \Delta t_{\mathrm{Radio}-\gamma}^{\mathrm{obs}}}{(1+z) \sin \theta}\\
\label{separation}
\end{equation}
where $\beta_{\mathrm{app}}$ is the apparent speed, $z$ is the redshift, $\theta$ is the viewing angle 
and the $\Delta t_{\mathrm{Radio}-\gamma}^{\mathrm{obs}}$ is the time lag in the observer's frame.
Using $\beta_{\mathrm{app}}$ $=$ 28~\citep{lister.21.apj}, $\theta$ $=$ 2.5$^\circ$~\citep{homam.21.apj} and $\Delta t_{\mathrm{Radio}-\gamma}^{\mathrm{obs}}$ $=$ 45 days which is the average of the time lags of the two segments with significant correlation, the distance between the two emission regions is estimated as 17.82 pc, and the corresponding projected distance is 0.78 pc.\\
\\
From the core shift measure ($\Omega$)~\citep[defined in][]{lobanov.98.aa} for PKS 1510$-$089 by~\citet{pushkarev.12.aa}, we can estimate the distance between the radio core and the true jet apex as 8.36 pc.
So, the value of the distance between the $\gamma$-ray emission region and the jet apex is a negative value.
Above results are in agreement with~\citet{kramarenko.22.mn} findings which calculated the radio lagging $\gamma$-ray emission $111_{-30}^{+16}$ days in the observer's frame and the $\gamma$-ray emission generation region is $-12.00_{-8.65}^{+13.71}$ pc from the central engine.
The jet is widely thought to be launched in the vicinity of the supermassive black hole within 100 $R_s$ (Schwarzschild radius)~\citep{meier.01.science}.
Referring to previous studies, the black hole mass of PKS 1510$-$089 is estimated to be on the order of $10^7$ to $10^8$ solar masses~\citep{oshlack.02.apj,xie.05.aj,liu.06.apj,abdo.10.apj,park.17.apj,castignani.17.aa,rakshit.20.aa}.
Therefore, the jet apex is $10^{-5}$ to $10^{-6}$ pc away from the central engine or supermassive black hole (SMBH), which is negligible.
The negative value of the distance from the location of the $\gamma$-ray emission production to the central engine might further suggest that the conjecture that the perturbation propagates in a straight jet geometry is too simple;
~\citet{kramarenko.22.mn} offer several alternative interpretations.\\

\subsubsection{Optical/NIR versus radio}
\noindent
 For this source, as discussed above, the optical and NIR emissions are dominated by non-thermal synchrotron emission from the relativistic jet and their emission region is almost co-spatial.
We see that the optical and radio ZDCF profiles are very similar to the NIR and radio ZDCF profiles, while the lack of some J-band data near MJD~57500 compared to the R-band data results in some visible difference between their profiles in segment 4.
Meanwhile, due to the seasonal gap in the data of this particular segment,
we cannot arrive at a firm conclusion about the correlation results in this segment.
However, the remaining three segments show that radio is lagging the optical/NIR radiation.
These results are in agreement with the above model;
in other words, the scenario in which radio emission lags behind optical/NIR emission is analogous to the scenario in which $\gamma$-ray emission lagging radio emission.
Although no firm conclusions have been drawn to date on the correlations between optical/NIR and radio emissions for blazars in general, correlation analyses beween optical/NIR and radio bands in large samples or individual sources have  revealed radio emission lagging the optical/NIR in a substantial fraction of cases \citep[e.g.][]{zhangbk.17.apjs,sarkar.19.apj}.

\subsection{QPO analysis}
\noindent
We found a hint of a possible QPO of about 1580 days in the 37 GHz radio emission, but it might only be confirmed through even longer future studies. 
Since the radio is synchrotron emission from the jet, jet-precession seems a potential interpretation of such a lengthy putative QPO. 
However, almost the entire multi-band emission is from the jet and thus, similar temporal profiles are expected across the entire EM bands unless there is a strong spectral change, which does not seem to be the case here.
Though the radio emission correlates very well with emission in other bands, the temporal profile is very different, with much smoother and broader peaks. 
Also, a large fraction of the radio photons often come from the unresolved core, as in the present case (see Fig.\ \ref{fig:mojave-core-total}). 

\section{Summary}\label{sec: conclusions}
\noindent
In this study, we performed a correlation analysis of the multi-band light curves of the FSRQ PKS 1510$-$089, for $\gamma$-ray, optical, NIR, and 37 GHz radio bands data sets spanning around ten years (February 2008 to September 2017).
The $\gamma$-ray data come from the Large Area Telescope onboard the space-based Fermi Gamma-ray Space Telescope, the optical R-band data and NIR J-band data are from the public archive of the SMARTS and the Steward Observatory telescopes, and the radio observations at 37~GHz were made at the 14~m radio telescope at the Mets{\"a}hovi Radio Observatory of Aalto University. 
We also employed several methods to search for the QPO phenomenon in the radio light curve between 2008 -- 2018. 
Through a series of analyses, we obtain the following results: \\
\\
$\bullet$ This source shows RWB behavior whether compared during individual segments or grouped by brightness states, indicating that jet emissions dominate accretion disk emissions throughout the observation period, even in low flux states. \\
$\bullet$ All segments show strong $\gamma$-ray and optical/NIR correlations,
and the correlation between $\gamma$-ray and radio is shown in the first two segments with a non-zero time lag. 
The existence of these correlations lends support to the lepton model of $\gamma$-ray emission. 
For PKS~1510$-$089, the nominal distance from the location of the $\gamma$-ray emission production to the central engine is a negative value using a simple model of shock propagating down a conical jet. 
This could indicate that the morphology of the conical jet may have been altered by jet dynamics, making the model unable to support such a complex situation. \\
$\bullet$ The correlation analysis reveals a clear, and unsurprising, correlation between the optical R-band and NIR J-band emissions, implying that these flares are simultaneous, or at least that the time lag between these light curves is shorter than the cadence of the observations.
Additionally, the optical and NIR flares are caused mainly by synchrotron radiation of the jet, and their emission regions are almost co-spatial. \\
$\bullet$ We confirm that variations in the optical/NIR band lead to variations in the 37 GHz radio band, and these results again suggest that the origin of the radiation that dominates these three frequencies is the same. 
This correlation indirectly indicates that other components, such as thermal radiation, make lesser contributions to the total optical and NIR fluxes.
This conclusion of jet dominance is further supported by the color index changes in the optical and NIR bands. \\
$\bullet$ A nominal QPO signal is found in the radio light curve, but the period of $\sim 1580$ d is too long compared to the length of the observations to be trusted.

\section*{Acknowledgments}

\noindent
This publication makes use of data obtained at Mets\"ahovi Radio Observatory, operated by Aalto University in Finland. 
The various diligent observers of Aalto University in Finland are thankfully acknowledged. 
This paper has made use of up-to-date SMARTS optical/near-infrared light curves that are available at \url{www.astro.yale.edu/smarts/glast/home.php}. 
SMARTS observations of Large Area Telescope-monitored blazars are supported by Yale University and Fermi GI grant NNX 12AP15G, and the SMARTS 1.3-m observing queue received support from NSF grant AST-0707627. 
Data from the Steward Observatory spectropolarimetric monitoring project were used. This programme is supported by Fermi GI grants NNX08AW56G, NNX09AU10G, NNX12AO93G and NNX15AU81G.
This work has made use of publicly available Fermi-LAT data obtained from FSSC’s website data server and provided by NASA Goddard Space Flight Center.
This research has made use of data from the MOJAVE database that is maintained by the MOJAVE team~\citep{lister.18.apjs}.\\ 
\\
This work is supported by the National Key R\&D Intergovernmental Cooperation Program of China (2022YFE0133700), the Regional Collaborative Innovation Project of Xinjiang Uyghur Autonomous Region (2022E01013), the National Natural Science Foundation of China (12173078, 11773062), and the Chinese Academy of Sciences (CAS) `Light of West China' Program (2017-XBQNXZ-A-008, 2021-XBQNXZ-005).
PK acknowledges support from the Department of Science and Technology (DST), government of India, through the DST-INSPIRE Faculty grant (DST/INSPIRE/04/2020/002586). 
ACG is partially supported by Chinese Academy of Sciences (CAS) President's International Fellowship Initiative (PIFI) (grant no.\ 2016VMB073). 
The work of AT was supported by the Innovation Program of the Shanghai Municipal Education Commission, Grant No. 2019-01-07-00-07-E00035, and the National Natural Science Foundation of China (NSFC), grant no.\ 11973019. 
MFG acknowledges support from the National Natural Science Foundation of China (grant no.\ 11873073), Shanghai Pilot Program for Basic Research Chinese Academy of Science, Shanghai Branch (JCYJ-SHFY2021-013), and the science research grants from the China Manned Space Project (No.\ CMSCSST-2021-A06).

\software{Python}

\bibliography{refs}{}
\bibliographystyle{aasjournal}

\end{document}